\documentclass[11pt,a4paper]{article}
\usepackage{jheppub}
\usepackage{lmodern}

\usepackage[T1]{fontenc}
\usepackage{units}
\usepackage{textcomp}
\usepackage{pmboxdraw}
\usepackage{mathrsfs}
\usepackage{amsmath}
\usepackage{amssymb}
\usepackage{graphicx}
\usepackage{epstopdf}
\usepackage{appendix}
\usepackage{setspace}
\usepackage{esint}

\makeatletter

\newcommand{\lyxmathsym}[1]{\ifmmode\begingroup\def\b@ld{bold}
  \text{\ifx\math@version\b@ld\bfseries\fi#1}\endgroup\else#1\fi}

\providecommand{\tabularnewline}{\\}

\@ifundefined{showcaptionsetup}{}{%
 \PassOptionsToPackage{caption=false}{subfig}}
\usepackage{subfig}
\makeatother
\preprint{FTUV-13-17-37, ~IFIC-13-45}
\title{Eviction of a 125 GeV ``heavy''-Higgs from the MSSM}
\author{G.~Barenboim}
\author{C.~Bosch}
\author{M.L.~L\'opez-Ib\'a\~nez}
\author{O. Vives}
\affiliation{Departament de F\'{\i}sica Te\`orica and IFIC, Universitat de 
Val\`encia-CSIC, E-46100, Burjassot, Spain.}

\abstract{
We prove that the present experimental constraints are already enough to rule out the possibility of the $\sim125$~GeV Higgs found at LHC being the second lightest Higgs in a general MSSM context, even with explicit CP violation in the Higgs potential. Contrary to previous studies, we are able to eliminate this possibility analytically, using simple expressions for a relatively small number of observables. We show that the present LHC constraints on the diphoton signal strength, $\tau\tau$ production through Higgs and BR($B \to X_s \gamma$) are enough to preclude the possibility of $H_2$ being the observed Higgs with $m_H\simeq 125$~GeV within an MSSM context, without leaving room for finely tuned cancellations.
As a by-product, we also comment on the difficulties of an MSSM interpretation of the excess in the $\gamma\gamma$ production cross section recently found at CMS that could correspond to a second Higgs resonance at $m_H\simeq 136$~GeV.}

\begin{document}

\maketitle

\section{Introduction}

In July 2012, both ATLAS and CMS, the two LHC general purpose experiments,
announced the discovery of a bosonic resonance with a mass 
$\sim125$~GeV that could be interpreted as the expected Higgs boson in the
Standard Model (SM) \cite{Aad:2012tfa,Chatrchyan:2012ufa}. 
The observed production cross section and decay channels
seem to be consistent, within errors, with a Higgs boson in the SM framework. 
However, at present, although CMS results are just below SM expectations, ATLAS  shows a slight excess in the most sensitive 
channels that, if confirmed
with more precise measurements, could be a sign of new physics beyond
the single SM Higgs. 

Besides, despite the extraordinary success of the SM in explaining all
the experimental results obtained so far, both in the high energy as
well as in the low energy region, there is a general belief that the
SM is not the ultimate theory, but only a low energy limit of a more
fundamental one.  This underling, more fundamental theory is expected
to contain new particles and interactions opening new processes not
possible in the SM but, above all, it is envisaged to go one step
further in the long way to reach a theory which incorporates gravity
to our quantum field description of Nature. In such an endeavor,
symmetries, who have historically played an important role in our
understanding of the laws of Nature, are expected to be a major
player.  This is one of the reasons why Supersymmetry (SUSY), the only
possible extension of symmetry beyond internal Lie symmetries and the
Poincare group \cite{Coleman:1967ad,Haag:1974qh}, is arguably the most popular extension of the SM. SUSY
is a symmetry between fermions and bosons, and, in its minimal
version, the Minimal Supersymmetric Standard Model (MSSM), assigns a
supersymmetric partner to each SM particle \cite{Fayet:1974pd,Fayet:1977yc,Farrar:1978xj,Witten:1981nf,Dimopoulos:1981zb,Sakai:1981gr,Ibanez:1981yh,Kaul:1981wp,Nilles:1983ge,Haber:1984rc}. These particles must have
a mass close to the electroweak scale, if SUSY is to solve the
hierarchy problem of the SM. Moreover, the MSSM requires a second
Higgs doublet in addition to the single doublet present in the SM and,
therefore, Higgs phenomenology in the MSSM is much richer than the SM,
with three neutral-Higgs states and a charged Higgs in the spectrum \cite{Djouadi:2005gj}.

At tree level, the scalar potential of the MSSM is CP-conserving, and
therefore mass eigenstates are also CP eigenstates. We have two
neutral scalar bosons, $h$ and $H$, and a neutral pseudoscalar,
$A$. However, the MSSM contains several CP violating phases beyond the
single SM phase in the CKM matrix\footnote{It is well-known that a single CKM
  phase is not enough to explain the observed matter-antimatter
  asymmetry of the universe. Additional phases (and therefore new
  physics) are required for that.}, {\it e.g.} $M_i, i=1,2,3$, $A_t$,
$\mu $ are complex parameters, and then CP violation necessarily leaks
into the Higgs sector at one-loop level \cite{Pilaftsis:1998dd,Pilaftsis:1998pe,Pilaftsis:1999qt,Demir:1999hj}. 
As a result, loop effects
involving the complex parameters in the Lagrangian violate the
tree-level CP-invariance of the MSSM Higgs potential modifying the
tree-level masses, couplings, production rates and decay widths of
Higgs bosons \cite{Pilaftsis:1999qt,Carena:2000yi,Choi:2000wz,Carena:2001fw,Choi:2001pg,Choi:2002zp}. In particular, the clear distinction between the two
CP-even and the one CP-odd neutral boson is lost and the
physical Higgs eigenstates become admixtures of CP-even and odd
states. Therefore, significant deviations from the naive CP conserving
scenario can be obtained in the regime where $M_{H^\pm}$ is low and
Im~$(\mu A_t)$ is significant. Yet, the size of SUSY phases is strongly constrained by searches of electric dipole moments (EDM) of the electron and neutron.  The phase of $\mu$ is bounded to be miserably small, $\lesssim 10^{-2}$, by the upper limits on EDMs if sfermion masses are below several TeV. Bounds on the phases of $A_{e,d,u}$, although somewhat weaker, are also strong, $\lesssim 10^{-1}$, under the same conditions. However, the phases of third generation trilinear couplings $A_{t,b,\tau}$ can still be sizeable\footnote{These phases enter EDMs of the electron and proton at two loops through Barr-Zee diagrams\cite{Barr:1990vd,Chang:1999zw}. However, these contributions are suppressed for heavy squarks\cite{Ellis:2008zy}.} for soft masses $O(1 ~\mbox{TeV})$ and, due to the large Yukawa couplings, these are precisely the couplings that influence the scalar potential more strongly \cite{Pilaftsis:1999td}. In this work, we will take only third-generation trilinear couplings  $A_{t,b,\tau}$ as complex to generate the scalar-pseudoscalar mixing in the Higgs potential. 

Among all the possibilities opened up by this scenario, 
one particularly interesting is the case where
the scalar observed at LHC is not the lightest but the second lightest 
one, having the lightest escaped detection at LEP/Tevatron/LHC due to its pseudoscalar or down-type content. 
As a result of the mixing, the couplings $H_1-WW$, $H_1-ZZ$ and $H_1-t\bar{t}$ 
all get reduced simultaneously evading the current bounds.
This idea of course is not new. Many studies have been carried out within
this model \cite{Heinemeyer:2011aa,Hagiwara:2012mga,Arbey:2012dq,Bechtle:2012jw,Ke:2012zq,Ke:2012yc,Moretti:2013lya,Scopel:2013bba}. There are two public codes, CPsuperH \cite{Lee:2003nta,Lee:2012wa}, specifically developed to analyze the Higgs phenomenology in the MSSM with explicit CP violation, and FeynHiggs \cite{Heinemeyer:1998yj,Hahn:2005cu}, that also calculates the spectrum and decay widths of the Higgses in the Complex MSSM. By using them, different regions of the parameters space have been 
explored through giant scans following the results of the colliders.

In this work, we will explore a different path. We will study this scenario,
not by scanning its parameters space but rather by choosing a pair of key
experimental signatures from both, high and low energy experiments, and 
analyzing (analytically or semi-analytically) whether their results can be simultaneously satisfied. This way we gain understanding on the physics
of the model we are discussing and at the same time avoid the possibility of
missing a fine-tuned region in the parameter space (even tiny to the point of being  microscopic) where an unexpected
cancellation or a lucky combination might occur. After all, whatever physics 
hides so effectively behind
the SM will turn out to be just one point in our studies of the parameter 
space. In this sense it is clear that every region, independently of
its size, has the same probability of being the right one and should
be given enough attention.

Moreover, our analysis is performed in terms of the SUSY parameters at
the electroweak scale, such that it encloses all possible MSSM setups
(including explicit CP violation), as the CMSSM, NUHM, pMSSM or even a
completely generic
MSSM\cite{Ellis:2002wv,Ellis:2002iu,Ellis:2008eu,Berger:2008cq,AbdusSalam:2009qd,Arbey:2012dq,Arbey:2012bp}. In
fact, only a handful of MSSM parameters affect the Higgs sector and
low-energy experiments that we study. As we will see, in the Higgs
sector, we fix $m_{H_1}\leq m_{H_2}\simeq 125~\mbox{GeV}\leq
m_{H_3}\simeq m_{H^\pm}\lesssim 200$--220~GeV and use the experimental
results to look for acceptable, $3\times3$, Higgs mixing matrices as a function of $\tan \beta$. Supersymmetric parameters affecting the Higgs sector, and also the indirect processes $B\to X_s \gamma$ and $B_s\to \mu^+ \mu^-$, are basically third generation masses and couplings, and gaugino masses. In our analysis, these parameters take general values consistent with the experimental constraints on direct and indirect searches.  

This paper is organized as follows. We begin by summarizing the experimental situation in Section~\ref{sec:experiment}. In Section~\ref{sec:model} we describe the basic ingredients of the 
model and analyze the direct and indirect signatures we will choose for our study. The parameter space is surveyed in Section~\ref{sec:analysis} and results and 
conclusions are contained in Section~\ref{sec:conclu}.

\section{Current experimental status.}
\label{sec:experiment}
\subsection{Higgs signal at the LHC.}

Both ATLAS and CMS experiments have recently updated the analysis
of the Higgs-like signal using the full $pp$ collision data sample.
The ATLAS analysis \cite{ATLAS-CONF-2013-034} uses
integrated luminosities of 4.8 fb$^{-1}$ at $\sqrt{s}=$7 TeV plus
20.7 fb$^{-1}$ at $\sqrt{s}=$8 TeV, for the most sensitive channels,
$H\rightarrow\gamma\gamma$, $H\rightarrow ZZ^{*}\rightarrow4l$ and
$H\rightarrow WW^{*}\rightarrow l\nu l\nu$, plus 4.7 fb$^{-1}$ at
$\sqrt{s}=$7 TeV and 13 fb$^{-1}$ at $\sqrt{s}=$8 TeV for the $H\rightarrow\tau\tau$
and $H\rightarrow b\bar{b}$. Similarly CMS study \cite{CMS-PAS-HIG-13-005}
uses 5.1 fb$^{-1}$ at $\sqrt{s}=$7 TeV and 19.8 fb$^{-1}$ at $\sqrt{s}=$8
TeV in all these channels.

The main channels contributing to the observed signal are the decays
into photons and two Z-bosons. On the other hand, the most relevant
channel constraining the presence of additional Higgs-bosons is the
decay into two $\tau$ leptons. ATLAS and CMS agree on the mass of
the observed state which is $m_{h}=124.3\pm0.6(\mbox{stat)\ensuremath{\pm}0.4(\mbox{sist)}}$~GeV
for ATLAS and $m_{h}=125.7\pm0.3(\mbox{stat})\pm 0.3(\mbox{sist})$~GeV
for CMS.

 However, there are some differences on the signal strength in the
different channels as measured by the two experiments. The signal
strength $\mu_{X}$, for a Higgs decaying to $X$ is defined as,
\begin{equation}
\mu_{X}=\frac{\sigma(pp\to H)\times\mbox{BR}(H\to X)}{\sigma(pp\to H)_{\rm{SM}}\times\mbox{BR}(H\to X)_{\rm{SM}}},
\end{equation}
\noindent such that $\mu=0$ corresponds to the background-only hypothesis
and $\mu=1$ corresponds to a SM Higgs signal. The combined signal
strength in the last results presented by ATLAS is $\mu^{\rm{ATLAS}}=1.3\pm0.2$ \cite{Aad:2013wqa},
while the signal strength measured by CMS is slightly below the SM
expectations $\mu^{\rm{CMS}}=0.80\pm0.14$ \cite{CMS-PAS-HIG-13-005}. 

For the diphoton channel, the measured signal strength in both experiments
are $\mu_{\gamma\gamma}^{\rm{ATLAS}}=1.6\pm0.3$ and $\mu_{\gamma\gamma}^{\rm{CMS}}=0.78_{-0.26}^{+0.28}$.
This signal is consistent with the SM, although ATLAS points to a
slight excess over the SM expectations. In any case, both results
agree on the fact that the diphoton signal must be of the order of
the SM prediction. This fact is very important in the context of multi-Higgs
models, as the MSSM, where the Higgs couplings to down quark and charged
leptons are enhanced by additional $\tan\beta$ factors, which tend
to decrease the $H\to\gamma\gamma$ branching ratio and therefore
the signal strength. In this regard, here we will adopt a conservative
approach and impose the weighted average of ATLAS and CMS results at 2$\sigma$,
\begin{equation}
0.75\leq\mu_{\gamma\gamma}^{\rm{LHC}}\leq1.55\,.
\end{equation}
\noindent Similarly, the signal strength in the $H\to ZZ^{*}$ channel
are, $\mu_{ZZ^{*}}^{\rm{ATLAS}}=1.5\pm0.4$ and $\mu_{ZZ^{*}}^{\rm{CMS}}=0.92\pm0.28$
and we will also use as a constraint, 
\begin{equation}
0.78\leq\mu_{ZZ^{*}}^{\rm{LHC}}\leq1.58\,.
\end{equation}

\begin{figure}
\subfloat[]{\includegraphics[scale=0.35]{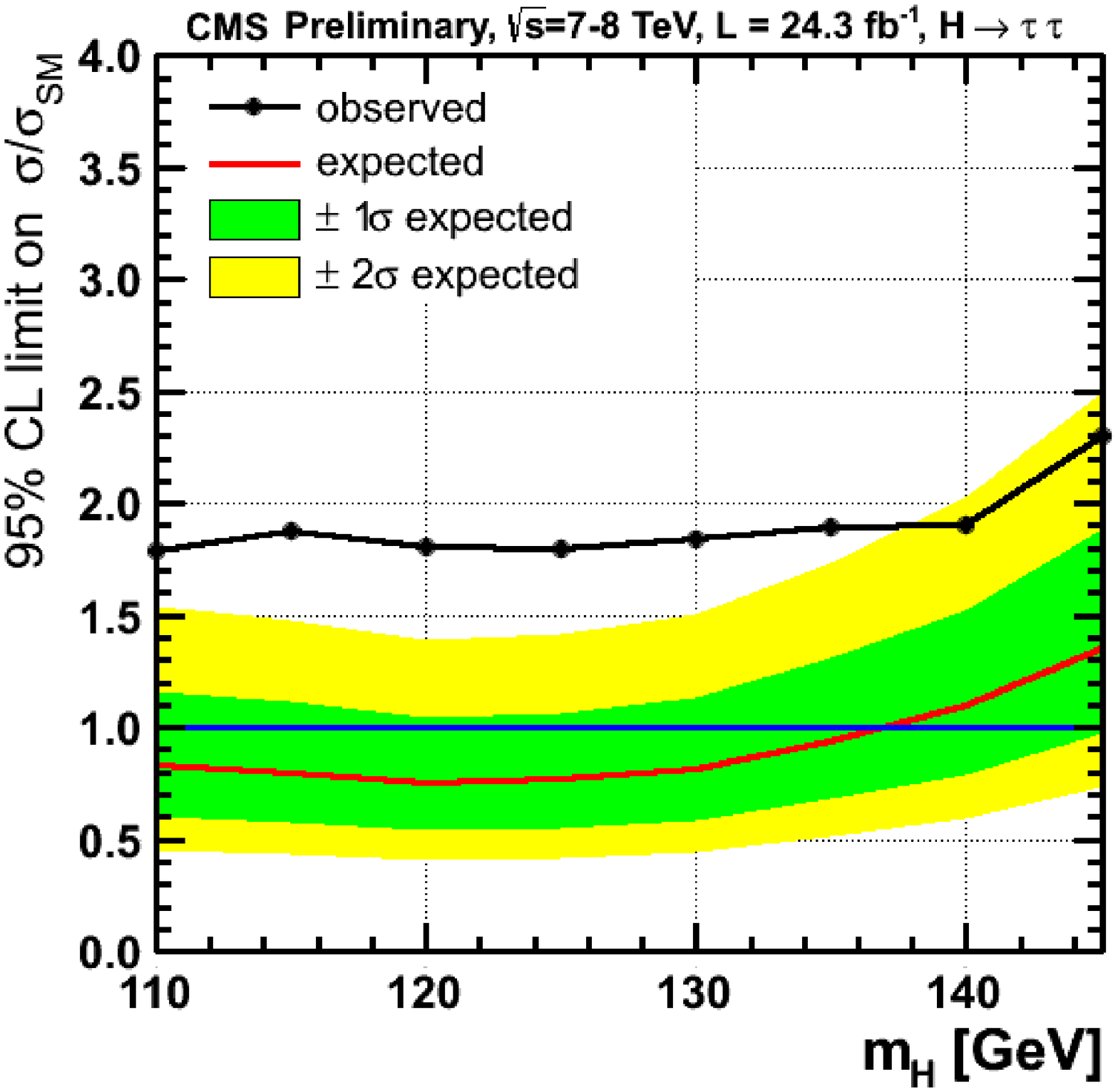}}\hspace{1cm}\subfloat[]{\includegraphics[scale=0.36]{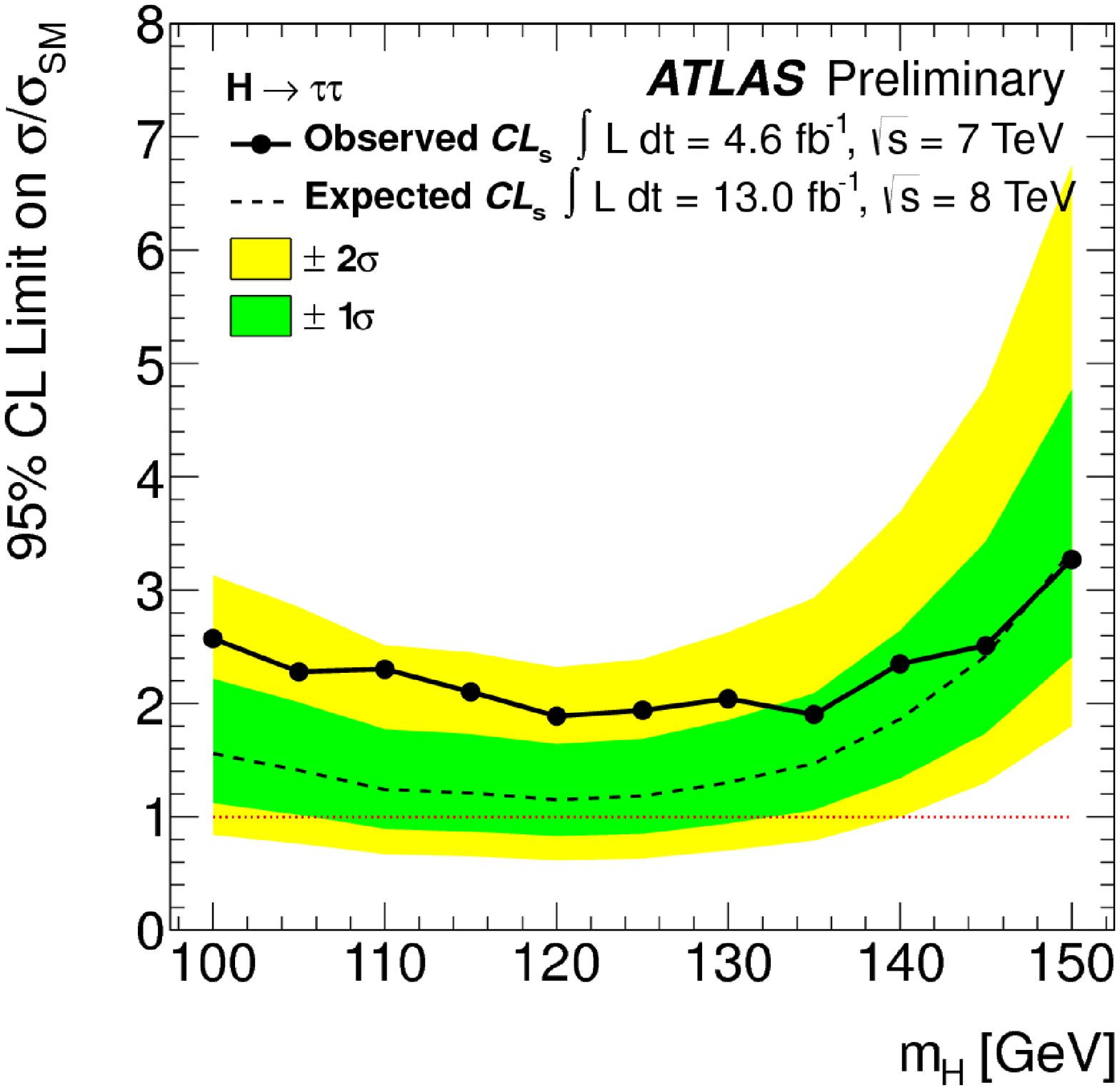}}\caption{\label{fig:tau-tauCERN}Higgs searches in the $H\rightarrow\tau\tau$ channel for $100~ \mbox{GeV}\leq m_H\leq 150~\mbox{GeV}$ at CMS (a) and ATLAS (b).}
\end{figure}

The main constraint on the presence of additional heavy Higgs states
comes from the $H/A\rightarrow\tau\tau$ searches at ATLAS and CMS experiments.
In this case, both experiments have searched for the SM Higgs boson
decaying into a pair of $\tau$-leptons and this provides a limit
on $\sigma(pp\to H)\times\mbox{BR}(H\to\tau\tau)$ that can be applied
to the extra Higgs states. ATLAS has analyzed the collected data samples
of $4.6\,\mbox{fb}^{-1}$at $\sqrt{s}=$7 TeV and $13.0\,\mbox{fb}^{-1}$at
$\sqrt{s}=$8 TeV \cite{Aad:2012mea} while CMS used $4.9\,\mbox{fb}^{-1}$at $\sqrt{s}=$7
TeV and $19.4\,\mbox{fb}^{-1}$at $\sqrt{s}=$8 TeV for Higgs masses
up to 150 GeV \cite{CMS-PAS-HIG-13-004}. These constraints on the $\tau\tau$-cross section
normalized to the SM cross section as a function of the Higgs mass
are shown in Figure \ref{fig:tau-tauCERN}. In this case, CMS sets
the strongest bound for $m_{H}$ below 150 GeV. For $m_{H}=110$ GeV
we obtain a bound at 95\% CL of $\mu_{\tau\tau}=\sigma\left(H\rightarrow\tau\tau\right)/\sigma_{SM}\leq1.8$, and this limit remains nearly constant, $\mu_{\tau\tau}\leq2.0$, up to $m_{H}=140$~GeV. For a neutral Higgs of mass $m_{H}=150$~GeV we would have a
bound of $\mu_{\tau\tau}\leq2.3$. In our scenario, this limit would apply to $H_1$ with a mass below 125 GeV and to $H_2$ with $m_{H_2}\simeq 125$~GeV. In the case of $H_3$, this bound applies for masses below 150 GeV.

\begin{figure}[t]
\noindent \begin{centering}
\includegraphics[scale=0.38]{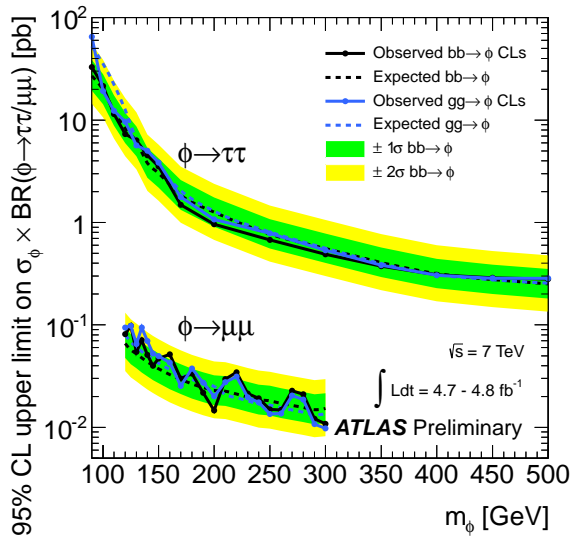}\caption{Upper limit on the $\tau\tau$ production cross section through heavy
Higgs states from ATLAS with $4.8~ \mbox{fb}^{-1}$ at $\sqrt{s}=7$~TeV \label{fig:ATLAS-MSSM-H}.}
\par\end{centering}
\end{figure}

For heavier $H_3$ masses, there exist a previous analysis at LHC searching
MSSM Higgs bosons with masses up to 500 GeV. In Figure \ref{fig:ATLAS-MSSM-H},
we present the analysis made in ATLAS with $4.9\,\mbox{fb}^{-1}$ at
$\sqrt{s}=$7 TeV \cite{Aad:2012yfa}. In this case, the bound is presented as an upper limit on the $\tau\tau$, or $\mu\mu$ production cross section. As a reference, the SM cross section for a Higgs mass of 150 GeV is $\sigma(pp\to H)_{\rm{SM}}\times\mbox{BR}(H\to X)_{\rm{SM}} \simeq 0.25$~pb and therefore, comparing with Figure~\ref{fig:tau-tauCERN}, we can expect this bound to improve nearly an order of magnitude in an updated analysis with the new data \cite{privateFiorini}. Nevertheless, the production cross-section of $\tau$-pairs through a heavy Higgs is enhanced by powers of $\tan\beta$ and therefore the present limits on $\sigma_{\phi}\times\mbox{BR}(\phi\to\tau\tau)$
are already very important in the medium--large $\tan\beta$ region.

Finally, we include the bounds on charged Higgs produced in $t \to H^+ b$ with subsequent decay $H^+ \to \tau \nu$ \cite{Aad:2012tj,CMS-PAS-HIG-12-052}. These analysis set upper bounds on $B(t \to H^+ b)$ in the range 2--3 \% for charged Higgs bosons with masses between 80 and 160 GeV, under the assumption that 
$B(H^+ \to \tau^+ \nu_\tau) = 1$, which is a very good assumption unless decay channels to the lighter Higgses and W-bosons are kinematically opened. 

\subsection{MSSM searches at LHC.}

Simultaneously to the Higgs searches described above, LHC has been looking for signatures
on new physics beyond the SM. A large effort has been devoted to search
for Supersymmetric extensions of the SM.
These studies, focused in searches of jets or leptons plus missing energy 
(possible evidence of the LSP), agree, so far, with the Standard Model expectations in all the explored region, and
are used to set bounds on the mass of the supersymmetric particles.

The most stringent constraints from LHC experiments are set on gluinos and first generation squarks produced through strong interactions in $pp$ collisions.
Searches of gluinos at CMS\cite{Chatrchyan:2012paa,Chatrchyan:2013wxa,PAS-SUS-13-007,PAS-SUS-13-008} and ATLAS \cite{ATLAS-CONF-2012-145,ATLAS-CONF-2013-007} with  $\sim20$~fb$^{-1}$ at 8 TeV have driven, roughly, to the exclusion of gluino masses up to 1.3 TeV for (neutralino) LSP masses below 500 GeV.
The limits on first generation squarks directly produced are $m_{\tilde q}\gtrsim 740$ GeV for squarks decaying $\tilde q \to q \chi_1^0$ with $m_{\chi_1^0}= 0$ GeV\cite{Chatrchyan:2013lya}\footnote{Limits on masses could be softer if these squarks are nearly degenerate with the LSP, but this does not affect our analysis below}.

The most important players in Higgs physics, because of their large Yukawa couplings, are third generation squarks. In this case mass bounds, from direct stop production, are somewhat weaker but still stop masses are required to be above $\sim 650$~GeV for $m_{\chi^0} \lesssim 200$~GeV \cite{ATLAS-CONF-2013-024,ATLAS-CONF-2013-037,ATLAS-CONF-2013-053,PAS-SUS-13-011} with the exception of small regions of nearly degenerate stop-neutralino.
Limits on sbottom mass from direct production are also similar and sbottom masses up to 620 GeV are excluded at 95\% C.L. for  $m_{\chi^0} < 150$~GeV, with the exception of  $m_{{\tilde b}_1}-m_{\chi^0}< 70$~GeV \cite{ATLAS-CONF-2013-053,Chatrchyan:2013lya,PAS-SUS-13-008}.

Finally, ATLAS and CMS have presented the limits on chargino masses from direct EW production \cite{ATLAS-CONF-2013-035,PAS-SUS-12-022}. In both analysis, these limits depend strongly on the slepton masses and the branching ratios of chargino and second neutralino that are supposed to be degenerate. When the decays to charged sleptons are dominant, chargino masses are excluded up to $\sim 600$~GeV for large mass differences with $\chi^0$. Even in the case when the slepton channels are closed, decays to weak bosons plus lightest neutralino can exclude\footnote{As pointed out in Ref.~\cite{Bharucha:2013epa}, these bounds with the slepton channel closed are only valid in a simplified model that assumes BR($\chi^0_2 \to Z \chi^0_1$)=1. This bound is strongly relaxed once the decay $\chi^0_2 \to h \chi^0_1$ is included. However, in our paper, this limit is only taken into account as a reference value for chargino masses and has no effect in our analysis of the feasibility of this scenario.} chargino masses up to $\sim 350$~GeV for $m_{\chi_1^0} \lesssim 120$~GeV. 

Therefore, as we have seen, limits on SUSY particles from LHC experiments are already very strong with the exceptions of sparticle masses rather degenerate with the lightest supersymmetric particle.  

\subsection{Indirect bounds}

Indirect probes of new physics in low energy experiments still play a
very relevant role in the search for extensions of the SM \cite{Masiero:2001ep,Raidal:2008jk,Calibbi:2011dn}. Even in the
absence of new flavour structures beyond the SM Yukawa couplings, in a
Minimal Flavour Violation scheme, decays like
$B_{s}^{0}\rightarrow\mu^{+}\mu^{-}$ and,
specially, $B\rightarrow X_{s}\gamma$ play a very important role,
as we will see below, and put significant constraints for the whole
$\tan\beta$ range.

The present experimental bounds on the decay $B_{s}^{0}\rightarrow\mu^{+}\mu^{-}$ are obtained from LHCb measurements with 1.1 fb$^{-1}$ of proton-proton collisions at $\sqrt{s} = 8$~TeV and 1.0 fb$^{-1}$ at $\sqrt{s} = 7$~TeV. The observed
value for the branching ratio at LHCb \cite{Aaij:2012nna,Aaij:2013aka} is,
\begin{equation}
 \mbox{BR}\left(B_{s}^{0}\rightarrow\mu^{+}\mu^{-}\right)=\left(2.9^{ +1.1}_{-1.0}
\right)\times10^{-9}\,,
\end{equation}  
and at CMS \cite{Chatrchyan:2013bka},
\begin{equation}
 \mbox{BR}\left(B_{s}^{0}\rightarrow\mu^{+}\mu^{-}\right)=\left(3.0^{ +1.0}_{-0.9}
\right)\times10^{-9}\,,
\end{equation} 
The limits on the decay $B\rightarrow X_{s}\gamma$ come from the BaBar and Belle  B-factories and CLEO \cite{Chen:2001fja,Abe:2001hk,Limosani:2009qg,Lees:2012wg,Lees:2012ufa,Aubert:2007my}. 
The current world average for $E_\gamma > 1.6$~GeV given by HFAG \cite{Amhis:2012bh,hfag} is,
\begin{equation}
  \mbox{BR}\left(B\rightarrow X_{s}\gamma\right)=\left(3.43\pm0.21\pm0.07\right)\times10^{-4}\,.
\end{equation}  
We will see that this result provides a very important constraint on the charged Higgs mass in the low $\tan \beta$ region where other supersymmetric contributions are small.

\section{Theoretical model}
\label{sec:model}
As explained in the introduction, we intend to investigate whether
the observed Higgs particle of $m_{H}\simeq125$~GeV could correspond
to the second Higgs in a general MSSM scenario, while the lightest
Higgs managed to evade the LEP searches \cite{Heinemeyer:2011aa,Hagiwara:2012mga,Arbey:2012dq,Bechtle:2012jw,Ke:2012zq,Ke:2012yc,Moretti:2013lya,Scopel:2013bba}. The scenario we consider
here is a generic MSSM defined at the electroweak scale. This means
we do not impose the usual mass relations obtained through RGE from
a high scale, that we obtain, for instance in the Constrained MSSM
(CMSSM), but keep all MSSM parameters as free and independent at $M_{W}$.
Furthermore, we are mainly interested in the Higgs sector of the model,
which we analyze assuming generic Higgs masses and mixings in the
presence of CP violation in the squark sector. 
\subsection{CP-violating MSSM Higgs sector}
As it is well-known, the Higgs sector of the MSSM consists of a type II two-Higgs
doublet model. In the MSSM, the scalar potential conserves CP at tree-level  \cite{Djouadi:2005gj}.
Nevertheless, in the presence of complex phases in the Lagrangian,
CP violation enters the Higgs potential at the one-loop level, resulting
in the mixing between the CP-even and CP-odd Higgses. Then, after
electroweak symmetry breaking, we have three physical neutral scalar
bosons, admixtures of the scalar and pseudoscalar Higgs bosons, plus
a charged Higgs boson \cite{Pilaftsis:1998dd,Pilaftsis:1998pe,Pilaftsis:1999qt,Demir:1999hj}. 

The Higgs fields in the electroweak vacuum, with vevs $\upsilon_{1}$and
$\upsilon_{2}$ and $\tan\beta=\upsilon_{2}/\upsilon_{1}$, are
\begin{equation}
\Phi_{1}=\left(\begin{array}{c}
\frac{1}{\sqrt{2}}\left(\upsilon_{1}+\phi_{1}+ia_{1}\right)\\\phi_{1}^{-}
\end{array}\right);\;\;\Phi_{2}=e^{i\xi}\left(\begin{array}{c}
\phi_{2}^{+}\\
\frac{1}{\sqrt{2}}\left(\upsilon_{2}+\phi_{2}+ia_{2}\right)
\end{array}\right)\,,\label{eq:3.1-2}
\end{equation}
and, as mentioned above, the presence of CP-violating phases in the
Lagrangian introduces off-diagonal mixing terms in the neutral Higgs
mass matrix. In the weak basis, $\left(\phi_{1},\phi_{2},a\right)$,
with $\phi_{1,2}$ CP-even, scalar, and $a=a_{1}\sin\beta+a_{2}\cos\beta$
the CP-odd, pseudoscalar state, we write the neutral Higgs mass matrix
as \cite{Pilaftsis:1999qt,Carena:2000yi,Carena:2001fw,Funakubo:2002yb},
\begin{equation}
M_{H}^{2}=\left(\begin{array}{cc}
M_{S}^{2} & M_{SP}^{2}\\
M_{PS}^{2} & M_{P}^{2}
\end{array}\right)\,,
\end{equation}
where the scalar-pseudoscalar mixings are non-vanishing in the presence
of phases, $M_{SP}^{2},M_{PS}^{2}\propto \mbox{Im}\left[\mu A_{t,b}e^{i\xi}\right]$.
Then, this $3\times3$ neutral Higgs mass matrix is diagonalized by
\begin{equation}
{\cal U}\cdot M_{H}^{2}\cdot{\cal U}^{T}=\mbox{Diag}\left(m_{H_{1}}^{2},m_{H_{2}}^{2},m_{H_{3}}^{2}\right)\,.
\end{equation}
The Higgs sector of the MSSM is defined at the electroweak scale at tree-level 
by only two parameters that, in the limit of
CP-conservation, are taken as $\left(m_{A}^{2},\tan\beta\right)$.
In the complex MSSM, the pseudoscalar Higgs is not a mass eigenstate
and its role as a parameter defining the Higgs sector is played by
the charged Higgs mass $m_{H^{\pm}}^{2}$. At higher orders, the different
MSSM particles enter in the Higgs masses and mixings, although the
main contributions are due to the top-stop and bottom--sbottom sectors. It is well-known that the one-loop corrections to $M_S^2$ can increase the lightest Higgs mass from $\lesssim M_Z$ to $\sim 130$~GeV \cite{Okada:1990vk,Ellis:1990nz,Haber:1990aw}, hence being $\lesssim M_Z$, with the leading part of order \cite{Haber:1996fp,Djouadi:2013vqa},
\begin{equation}
\delta M_S^2 \simeq \frac{3 m_t^4}{2 \pi^2 \upsilon^2 \sin^2\beta} \left[\log \frac{M_{SUSY}^2}{ m_t^2} + \frac{X_t^2}{M_{SUSY}^2} \left(1- \frac{X_t^2}{12 M_{SUSY}^2}\right)\right]\,,
\end{equation}
with $M_{SUSY}$ the geometric mean of the two stop masses and  $X_t = A_t -\mu \cot \beta$. 
 
Regarding the charged Higgs mass, we can relate it to the pseudoscalar mass $M_{P}^{2}$ in the neutral Higgs mass matrix \cite{Pilaftsis:1999qt},
\begin{equation}
M_{H^{\pm}}^{2}=M_{P}^{2}+\frac{1}{2}\lambda_{4}\upsilon^{2}-\mbox{Re}\left(\lambda_{5}e^{2i\xi}\right)\upsilon^{2}\,,
\end{equation}
 with $\lambda_{4,5}$ the two-loop corrected parameters of the Higgs
potential \cite{Carena:1995bx,Pilaftsis:1999qt}. At tree level $\lambda_{4}=g_{w}^{2}/2$,
such that $\lambda_{4}\upsilon^{2}/2=M_{W}^{2}$, and $\lambda_{5}=$0.
In any case, it looks reasonable to expected $\lambda_{i}\lesssim1$. This implies that
the squared charged Higgs mass can never be heavier that the largest
neutral Higgs eigenvalue by a difference much larger than $M_{Z}^{2}$, which is equivalent to say that loop corrections are of the same order as 
$\sim\delta M_S^2$.

Similarly, we can expect the mass of the second neutral Higgs, which in our scenario is $m_{H_{2}}\simeq125$~GeV, only to
differ from the heavier eigenvalue by terms of order $\upsilon^{2}$. 
This can be seen from the trace of the neutral Higgs masses in the basis of CP eigenstates, where we would have, without loop corrections, $\mbox{Tr}\left(M_{H}^{2}\right)=2M_{P}^{2}+M_{Z}^{2}$. As we have seen, loop corrections to the diagonal elements can be expected to be of the order of the corrections to the lightest Higgs mass which are also $O(M_Z^2)$. To obtain a light second Higgs we need, either low $M_{P}$  or a large scalar-pseudoscalar mixing. The different contributions to scalar-pseudoscalar mixing, $M_{SP}^2$, are of order \cite{Pilaftsis:1999qt},
\begin{equation}
M_{SP}^2 = O\left(\frac{m_t^4 |\mu| |A_t|}{32 \pi^2~\upsilon^2 M_{SUSY}^2}\right) \sin\phi_{CP} \times \left[6, \frac{|A_t|^2}{M_{SUSY}^2}, \frac{|\mu|^2}{\tan\beta M_{SUSY}^2}\right]\,,
\end{equation}
which again are of the same order as $\delta M_S^2\simeq O(M_Z^2)$ for $\sin \phi_{CP} \sim O(1)$. Therefore, taking also into account that in the decoupling limit, and in the absence of scalar-pseudoscalar mixing, $M_H\simeq M_P$, we must require $M_{P}^{2}$ not to be much larger than $M_Z^2$. Taking $M_{P}^{2}\lesssim3M_{Z}^{2}$,  the invariance of the trace 
tells us that $m_{H_{1}}^{2}+m_{H_{2}}^{2}+m_{H_{3}}^{2}=2M_{P}^{2}+M_{Z}^{2}+ O(M_Z^2)$ in such a way that with $90~\mbox{GeV}\lesssim m_{H_{1}}\lesssim m_{H_{2}}\simeq125$~GeV, we get an upper limit\footnote{Allowing the heaviest neutral Higgs to be $200$~GeV with a second-heaviest Higgs of 125~GeV is a very conservative assumption. However, it looks very difficult to have such a heavy Higgs in any realistic MSSM construction.} for $m_{H_{3}}^{3}\lesssim 2M_{P}^{2}+ 2 M_{Z}^{2}- \left(m_{H_{2}}^{2}+ m_{H_{1}}^2\right) \lesssim(200~\mbox{GeV})^2$.
We must emphasize that in this work we do not consider the possibility of $m_{H_{1}}\lesssim 90~\mbox{GeV}$ which would be possible in the presence of large CP-violating phases that could reduce the mass of the lightest Higgs through rather precise cancellations \cite{Carena:2000ks,Carena:2002bb}. Although this scenario could survive LEP limits around an ``open hole'' with $m_{H_{1}}\approx 45~\mbox{GeV}$ and $\tan \beta \approx 8$ \cite{Williams:2007dc}, it would never be able to reproduce the observed signal in $H_2 \to \gamma \gamma$, as the opening of the decay channel $H_2\to H_1 H_1$ would render $B(H_2 \to \gamma \gamma)$ much smaller than the SM one (see the discussion related to the $H_2 \to b \bar{b}$ channel below).     
 
In the following analysis of the direct and indirect constraints on
the Higgs sector, we try to be completely general in the framework
of a Complex MSSM defined at the electroweak scale. To attain this
objective, and taking into account that the presence of CP violation
and large radiative corrections strongly modifies the neutral Higgs
mass matrix if we are outside the decoupling regime, we consider general neutral Higgs mixings and masses.
In fact, in this work, we analyze the situation in which the second
lightest neutral boson corresponds to the scalar resonance measured
at LHC with a mass of 125 GeV. As we have seen, to achieve this, we
need a relatively light charged Higgs (with approximately $M_{H^{+}}\lesssim220\,$~GeV), and a similar mass for the
heaviest neutral Higgs. The lightest neutral Higgs boson will have
a mass varying in the range of 90 and 125~GeV. After fixing the Higgs
masses in these ranges, we will consider generic mixing matrices ${\cal U}$
and look for mixings consistent with the present experimental results.

This analysis deals with the decays of the neutral Higgs bosons. Thus
we need the Higgs couplings to the SM vector boson, fermions, scalars
and gauginos. The conventions used in the following are described in 
Appendix~\ref{App:convention}. The couplings to the vector bosons are \cite{Lee:2003nta},
\begin{equation}
{\cal L}_{H_{a}V}=g\, M_{W}\left(W_{\mu}^{+}W^{-\,\mu}+\frac{1}{2\cos^{2}\theta_{W}}Z_{\mu}Z^{\mu}\right)\sum_{a}g_{H_{a}VV}\, H_{a}\,.
\end{equation}
with $g_{H_{a}VV}=\cos\beta\,\mathcal{U}_{a1}+\sin\beta\,\mathcal{U}_{a2}$.

The Lagrangian showing the fermion--Higgs couplings is  
\begin{equation}
{\cal L}_{H_{a}f}=-\sum_{f}\frac{g\, m_{f}}{2M_{W}}\sum_{a}H_{a}\bar{f}\left(g_{S,a}^{f}+ig_{P,a}^{f}\gamma_{5}\right)f\,,
\end{equation}
\begin{table}
\begin{centering}
{\Large \begin{tabular}{|c|c|c|c|}
\hline 
{ $\mathbf{H_{a}\rightarrow f\bar{f}}$} & { $\mathbf{g_{f}}$} & { $\mathbf{g_{S,a}^{(0)}}$} & { $\mathbf{g_{P,a}^{(0)}}$}\tabularnewline
\hline 
{ $H_{a}\rightarrow l\bar{l}$} & { $\frac{gm_{l}}{2M_{W}}$} & { $\frac{U_{a1}}{\cos(\beta)}$} & { $-\left(\frac{\sin(\beta)}{\cos(\beta)}\right)U_{a3}$}\tabularnewline
\hline 
{ $H_{a}\rightarrow d\bar{d}$} & { $\frac{gm_{d}}{2M_{W}}$} & { $\frac{U_{a1}}{\cos(\beta)}$} & { $-\left(\frac{\sin(\beta)}{\cos(\beta)}\right)U_{a3}$}\tabularnewline
\hline 
{ $H_{a}\rightarrow u\bar{u}$} & { $\frac{gm_{u}}{2M_{W}}$} & { $\frac{U_{a2}}{\sin(\beta)}$} & { $-\left(\frac{\cos(\beta)}{\sin(\beta)}\right)U_{a3}$}\tabularnewline
\hline 
{ $H_{a}\rightarrow\tilde{\chi}_{i}^{+}\tilde{\chi}_{j}^{-}$} & { $\frac{g}{\sqrt{2}}$} & { $g_{s}^{\tilde{\chi}^{+}}$} & { $g_{p}^{\tilde{\chi}^{+}}$}\tabularnewline
\hline 
\end{tabular}}
\par\end{centering}
\caption{Tree level Higgs--fermion couplings.}
\label{tab:Hfcoupl}
\end{table}
where the tree-level values of $(g_{S}^{(0)},g_{P}^{(0)})$ are given
in Table~\ref{tab:Hfcoupl}. Still, in the case of third generation fermions, these
couplings receive very important threshold corrections due to gluino
and chargino loops enhanced by $\tan\beta$ factors in the case of
the down-type fermions \cite{Hall:1993gn,Carena:1994bv,Blazek:1995nv,Carena:1999py,Hamzaoui:1998nu,Babu:1999hn,Isidori:2001fv,Dedes:2002er,Buras:2002vd}. The complete corrected couplings for third generation
fermions, $(g_{S}^{f},g_{P}^{f})$, can be found in Ref.~\cite{Lee:2003nta,Carena:2002bb}. In our
analysis, it is sufficient to consider the correction to the bottom
couplings, 
\begin{equation}
\label{Eq:thresholdS}
g_{S,a}^{d}=\mbox{Re}\left(\frac{1}{1+\kappa_{d}\tan\beta}\right)\frac{{\cal U}_{a1}}{\cos\beta}+\mbox{Re}\left(\frac{\kappa_{d}}{1+\kappa_{d}\tan\beta}\right)\frac{{\cal U}_{a2}}{\cos\beta}+\mbox{Im}\left(\frac{\kappa_{d}\left(\tan^{2}\beta+1\right)}{1+\kappa_{d}\tan\beta}\right){\cal U}_{a3}
\end{equation}
\begin{equation}
\label{Eq:thresholdP}
g_{P,a}^{d}=-\mbox{Re}\left(\frac{\tan\beta-\kappa_{d}}{1+\kappa_{d}\tan\beta}\right){\cal U}_{a3}+\mbox{Im}\left(\frac{\kappa_{d}\tan\beta}{1+\kappa_{d}\tan\beta}\right)\frac{{\cal U}_{a1}}{\cos\beta}-\mbox{Im}\left(\frac{\kappa_{d}}{1+\kappa_{d}\tan\beta}\right)\frac{{\cal U}_{a2}}{\cos\beta}
\end{equation}
where $\kappa_{d}=(\Delta h_{d}/h_{d})/(1+\delta h_{d}/h_{d})$ and
the corrected Yukawa couplings are,
\begin{equation}
h_{d}=\frac{\sqrt{2}m_{d}}{\upsilon\cos\beta}\:\frac{1}{1+\delta h_{d}/h_{d}+\Delta h_{d}/h_{d}\tan\beta}\,,
\end{equation}
\begin{eqnarray}
\delta h_{d}/h_{d}&=&-\frac{2\alpha_{s}}{3\pi}m_{\tilde{g}}^{*}A_{d}\, I(m_{\tilde{d}_{1}}^{2},m_{\tilde{d}_{2}}^{2},|m_{\tilde{g}}|^{2})-\frac{|h_{u}|^{2}}{16\pi^{2}}|\mu|^{2}\, I(m_{\tilde{u}_{1}}^{2},m_{\tilde{u}_{2}}^{2},|\mu|^{2}) \nonumber\\
\Delta h_{d}/h_{d}&=&\frac{2\alpha_{s}}{3\pi}m_{\tilde{g}}^{*}\mu^{*}\, I(m_{\tilde{d}_{1}}^{2},m_{\tilde{d}_{2}}^{2},|m_{\tilde{g}}|^{2})+\frac{|h_{u}|^{2}}{16\pi^{2}}A_{u}^{*}\mu^{*}\, I(m_{\tilde{u}_{1}}^{2},m_{\tilde{u}_{2}}^{2},|\mu|^{2})\:,
\end{eqnarray}
and the loop function $I(a,b,c)$ is given by,
\begin{equation}
I(a,b,c)=\frac{a\, b\log(a/b)+b\, c\log(b/c)+a\, c\log(c/a)}{(a-b)(b-c)(a-c)}\,.
\end{equation}
The Higgs-sfermion couplings are,
\begin{eqnarray}
{\cal L}_{H_{a}\tilde f \tilde f}= \upsilon \sum_{\tilde f} g_{\tilde f\tilde f }^a  \left(H_{a}\tilde{f}^* \tilde{f}\right)\,,
\\
 \upsilon~ g_{\tilde f_i\tilde f_j }^a = \left(\tilde{\Gamma}^{\alpha f f}\right)_{\beta \gamma} {\cal U}_{a \alpha}~ {\cal R}^f_{\beta i} {\cal R}^f_{\gamma j}\,,
\end{eqnarray}
with $\beta,\gamma = L,R$, $ {\cal R}^f$, the sfermion mixing matrices and the couplings $\tilde{\Gamma}^{\alpha f f}$ given Ref.~\cite{Lee:2003nta}.
Other Higgs couplings that are needed to analyze the neutral Higgs
decays are the couplings to charginos and charged Higgs,
complete expressions can be found in Ref.~\cite{Lee:2003nta} (taking into account their different convention on the Higgs mixing matrix, ${\cal U}={\cal O}^{T}$).

After defining all these couplings, we show in the following the expressions
for $H\to\gamma\gamma$
and $H\to gg$,  that together with $H\rightarrow\bar{b}b,\tau\tau$ and $H\to WW^{*},ZZ^{*}$ are the main Higgs decay channels for $m_{H}=125$~GeV, and the Higgs production mechanisms at LHC.

\subsection{Higgs decays.}
\subsubsection{Higgs decay into two photons.}
The decay $H_{a}\to\gamma\gamma$ occurs only at the one-loop level
and therefore we must include every contribution generated by sparticles
in addition to the SM ones in our calculation. Taking into account the presence
of CP violation, the Higgs decay has contributions of both the scalar
and pseudoscalar components. Then its width becomes, 
\begin{equation}
\Gamma\left(H_{a}\rightarrow\gamma\gamma\right)=\frac{M_{H_{a}}^{3}\alpha^{2}}{256\pi^{3}\upsilon^{2}}\left[\left|S_{a}^{\gamma}\left(M_{H_{a}}\right)\right|^{2}+\left|P_{a}^{\gamma}\left(M_{H_{a}}\right)\right|^{2}\right]\,,\label{eq:3.3.2-5-1}
\end{equation}
where the scalar part is $S_{a}^{\gamma}\left(M_{H_{a}}\right)$ and
the pseudoscalar $P_{a}^{\gamma}\left(M_{H_{a}}\right)$ and they are \cite{Lee:2003nta},
\begin{eqnarray}
S_{a}^{\gamma}\left(M_{H_{a}}\right)&=& 2\underset{f=b,t,\tilde{\chi}_{1}^{\pm},\tilde{\chi}_{2}^{\pm}}{\sum}N_{C}\, J_{f}^{\gamma}\, Q_{f}^{2}g_{f}\, g_{H_{a}\bar{f}f}^{S}\frac{\upsilon}{m_{f}}F_{f}^{S}\left(\tau_{af}\right) -\sum_{\tilde f} N_{C}\, J_{\tilde{f}}^{\gamma}\, Q_{f}^{2}\, g_{H_{a}\tilde{f}_{j}\tilde{f_{j}^*}}^{S}\frac{\upsilon^{2}}{2m_{\tilde{f}_{j}}^{2}}F_{0}\left(\tau_{a\tilde{f}_{j}}\right)\nonumber \\
 &&-g_{H_{a}VV}F_{1}\left(\tau_{aW}\right)~-~g_{H_{a}H^{-}H^{+}}\frac{\upsilon^{2}}{2M_{H_{a}}^{2}}F_{0}\left(\tau_{aH}\right)\\
P_{a}^{\gamma}\left(M_{H_{a}}\right)&= & 2\underset{f=b,t,\tilde{\chi}_{1}^{\pm},\tilde{\chi}_{2}^{\pm}}{\sum}N_{C}\, J_{f}^{\gamma}\, Q_{f}^{2}g_{f}\, g_{H_{a}\bar{f}f}^{P}\frac{\upsilon}{m_{f}}F_{f}^{P}\left(\tau_{af}\right)
\label{eq:3.3.2-2}
\end{eqnarray}
With $\tau_{aj}=M_{H_{a}}^{2}/(4m_{i}^{2})$ and the loop functions
being:
\begin{equation}
\begin{array}{ll}
F_{f}^{S}\left(\tau\right)=\tau^{-1}\left[1+\left(1-\tau^{-1}\right)f\left(\tau\right)\right];\qquad & F_{f}^{P}\left(\tau\right)=\tau^{-1}f\left(\tau\right);\\
F_{0}\left(\tau\right)=\tau^{-1}\left[-1+\tau^{-1}f\left(\tau\right)\right]; & F_{1}\left(\tau\right)=2+3\tau^{-1}+3\tau^{-1}\left(2-\tau^{-1}\right)f\left(\tau\right);
\end{array}\label{eq:3.3.2-3}
\end{equation}
\begin{equation}
f\left(\tau\right)=-\frac{1}{2}\intop\nolimits _{0}^{1}\frac{\mathrm{d}x}{x}\ln\left[1-4\tau x\left(1-x\right)\right]=\begin{cases}
\arcsin^{2}\left(\sqrt{\tau}\right)\quad: & \tau\leq1\\
\\
-\frac{1}{4}\left[\ln\left(\frac{\sqrt{\tau}+\sqrt{\tau-1}}{\sqrt{\tau}-\sqrt{\tau-1}}\right)-i\pi\right]^{2}\quad: & \tau\geq1
\end{cases}\label{eq:3.3.2-4}
\end{equation}
And we included the QCD corrections \cite{Spira:1995rr,Spira:1997dg},
\begin{equation}
J_{\chi}^{\gamma}=1;\; J_{q}^{\gamma}=1-\frac{\alpha_{s}\left(M_{H_{a}}^{2}\right)}{\pi};\qquad J_{\tilde{q}}^{\gamma}=1+\frac{\alpha_{s}\left(M_{H_{a}}^{2}\right)}{\pi}\label{eq:3.3.2-6}
\end{equation}
\subsubsection{Higgs decay into two gluons.}
Similarly, the decay width for $H_{a}\rightarrow gg$ is given by:
\begin{equation}
\Gamma_{H_{a}\rightarrow gg}=\frac{M_{H_{a}}^{2}\alpha_{s}^{2}}{32\pi^{3}v^{2}}\left[K_{H}^{g}|S_{a}^{g}|^{2}+K_{A}^{g}|P_{a}^{g}|^{2}\right]\label{eq:3.3.3-1}
\end{equation}
where $K_{H,A}^{g}$ is again the QCD correction enhancement factor
while $S_{a}^{g}$ and $P_{a}^{g}$ are the scalar and pseudoscalar
form factors, respectively. $K_{H,A}^{g}$ is \cite{Spira:1995rr,Spira:1997dg},
\begin{equation}
K_{H}^{g}=1+\frac{\alpha_{s}(M_{H_{a}}^{2})}{\pi}\left(\frac{95}{4}-\frac{7}{6}N^{F}\right)\, ,\qquad
K_{A}^{g}=1+\frac{\alpha_{s}(M_{H_{a}}^{2})}{\pi}\left(\frac{97}{4}-\frac{7}{6}N^{F}\right)\,,\label{eq:3.3.3-3}
\end{equation}
being $N^{F}$ the number of quark flavours that remains lighter than
the Higgs boson in consideration. On the other hand, the expressions
that define $S_{a}^{g}$ and $P_{a}^{g}$ are:
\begin{equation}
S_{a}^{g}=\sum_{f=b,t}g_{f}\, g_{sff}^{a}\frac{v}{m_{f}}F_{f}^{S}(\tau_{af})~-\sum_{\bar{f}_{i}=\tilde{b}_{1},\tilde{b}_{2},\tilde{t}_{1},\tilde{t}_{2}}g_{\tilde{f}\tilde{f}}^{a}\frac{v^{2}}{4m_{\bar{f}_{i}}^{2}}F_{0}(\tau_{a\tilde{f}_{i}})\label{eq:3.3.3-4}
\end{equation}
\begin{equation}
P_{a}^{g}=\sum_{f=b,t}g_{f}\, g_{pff}^{a}\frac{v}{m_{f}}F_{f}^{P}(\tau_{af})\label{eq:3.3.3-5}
\end{equation}
\subsection{Higgs production.\label{sub:Higgs-production.}}
The Higgs production processes are basically the same as in the SM \cite{Djouadi:2005gi,Djouadi:2005gj},
although the couplings in these processes change to the MSSM couplings.
The two main production processes are gluon fusion and, specially
for large $\tan\beta$, the $b\bar{b}$ fusion. Other production mechanisms,
like vector boson fusion will always be sub-dominant and we do not
consider them here.

At parton level, the leading order cross section for the production
of Higgs particles through the gluon fusion process is given by \cite{Dedes:1999sj,Dedes:1999zh,Choi:1999aj,Djouadi:2005gj}:
\begin{eqnarray}
\sigma_{gg\rightarrow H_{a}}^{LO}&=&\hat{\sigma}_{gg\rightarrow H_{a}}^{LO}\:\delta\left(1- \frac{M_{H_{a}}^{2}}{\hat{s}}\right)=\frac{\pi^{2}}{8M_{H_a}}\Gamma_{H_{a}\to gg}^{LO}\:\delta\left(1-\frac{M_{H_{a}}^{2}}{\hat{s}}\right)\\
\hat{\sigma}_{gg\rightarrow H_{a}}^{LO} & =&\frac{\alpha_{s}^{2}\left(Q\right)}{256\pi}\frac{M_{H_a}^2}{\upsilon^2}  \left[\left|\sum_{f=t,b}\frac{g_f g_{S,a}^{f} \upsilon}{m_{f}}F_{f}^{S}\left(\tau_{af}\right)+\frac{1}{4}\sum_{\tilde{f}_{i}=\tilde{b}_{1},\tilde{b}_{2},\tilde{t}_{1},\tilde{t}_{2}}\frac{g_{\tilde{f}\tilde{f}}^{a}\upsilon^2}{m_{\tilde{f}}^{2}}F_{0}\left(\tau_{a\tilde{f}}\right)\right|^{2}\nonumber\right. \\& +&\left. \left|\sum_{f=t,b}\frac{g_f g_{P,a}^{f}\upsilon}{m_{f}}F_{f}^{P}\left(\tau_{af}\right)\right|^{2}\right]
~~=~\frac{\alpha_{s}^{2}\left(Q\right)}{256\pi}\frac{M_{H_a}^2}{\upsilon^2} \Bigg[\left|S^g_a\right|^2+\left|P^g_a \right|^{2}\Bigg]\,,\label{eq:3.3.4-2}\nonumber
\end{eqnarray}
with $\hat{s}$ the partonic center of mass energy squared. The hadronic
cross section from gluon fusion processes can be obtained in the narrow-width
approximation as,
\begin{equation}
\sigma(pp\to H_{a})^{LO}=\hat{\sigma}_{gg\rightarrow H_{a}}^{LO}\tau_{H_{a}}\frac{d{\cal L}_{LO}^{gg}}{d\tau_{H_{a}}}\:.
\end{equation}
The gluon luminosity $d{\cal L}_{LO}^{gg}/d\tau$ at the factorization
scale $M$, with $\tau_{H_a}=M_{H_a}^2/s$, is given by, 
\begin{equation}
\frac{d{\cal L}_{LO}^{gg}}{d\tau}=\int_{\tau}^{1}\frac{dx}{x}\, g(x,M^{2})\, g(\tau/x,M^{2})\,.
\end{equation}
In the numerical analysis below, we use the MSTW2008 \cite{Martin:2009iq} parton distribution functions.

The $bb\rightarrow H_{a}$ production process can also play an important
role for the high and intermediate $\tan\beta$ region, roughly for
$\tan\beta\geq7$ \cite{Dicus:1988cx,Campbell:2002zm,Maltoni:2003pn,Harlander:2003ai,Dittmaier:2003ej,Dawson:2003kb,Baglio:2010ae}. The leading order partonic cross section is directly
related to the fermionic decay width,
\begin{eqnarray}
\hat{\sigma}_{bb\rightarrow H_{a}} & = & \frac{4\pi^{2}}{9M_{H_{a}}}\Gamma_{H_{a}\rightarrow b\bar{b}}=  \frac{\pi}{6}\frac{g^{2}m_{b}^{2}}{4M_{W}^{2}}\beta_{b}\left(\beta_{b}^{2}\left|g_{s}^{b}\right|^{2}+\left|g_{p}^{b}\right|^{2}\right)\label{eq:3.3.5-1}
\end{eqnarray}
Again the proton-proton cross section is obtained in the narrow-width
approximation in terms of the $b\bar{b}$ luminosity. Notice that
associated Higgs production with heavy quarks $gg/q\bar{q}\to b\bar{b}+H_{a}$
is equivalent to the $b\bar{b}\to H_{a}$ inclusive process if we
do not require to observe the final state $b$-jets and one considers
the $b$-quark as a massless parton in a five active flavour scheme
\cite{Dicus:1988cx,Djouadi:2005gj,Djouadi:2013vqa}. In this way, large logarithms $\log(s/m_b^2)$ are resummed to all orders.
As before, we are using the MSTW2008 five flavour parton distribution functions.
Regarding the QCD corrections to this process, for our purposes it is enough to take into account the QCD enhancing factor $K^f_a$ used in the decay $H_a\to b\bar b$, with the bottom mass evaluated at $m_{H_a}$, and to use the threshold-corrected bottom couplings in Eqs.~(\ref{Eq:thresholdS},\ref{Eq:thresholdP}).
\begin{eqnarray}
\hat{\sigma}_{bb\rightarrow H_{a}}^{QCD} & = & \frac{4\pi^{2}}{9M_{H_{a}}}\Gamma_{H_{a}\rightarrow b\bar{b}}=  \frac{\pi}{6}\frac{g^{2}m_{b}^{2}}{4M_{W}^{2}}K_{a}^{b}\left(\frac{m_{b}(m_{H_{a}})}{m_{b}(m_{t})}\right)^{2}\beta_{b}\left(\beta_{b}^{2}\left|g_{s}^{b}\right|^{2}+\left|g_{p}^{b}\right|^{2}\right)\label{eq:bbfus}
\end{eqnarray}
The total hadronic cross section can be obtained at NLO using the so-called $K$-factors \cite{Spira:1997dg,Graudenz:1992pv,Dawson:1996xz,Choi:1999aj} to correct the LO gluon fusion, and it is given by,
\begin{equation}
\sigma(pp\to H_{a})=K~\hat{\sigma}_{gg\rightarrow H_{a}}^{LO}\tau_{H_{a}}\frac{d{\cal L}_{LO}^{gg}}{d\tau_{H_{a}}}\:+\:\hat{\sigma}_{bb\rightarrow H_{a}}^{QCD}\tau_{H_{a}}\frac{d{\cal L}_{LO}^{bb}}{d\tau_{H_{a}}}
\end{equation}
where the $K$-factor parametrizes the ratio of the higher order cross
section to the leading order one. It is  important to include
this term as it is known that the next to 
leading order QCD effects, which affect both quark and squark contributions similarly \cite{Dawson:1996xz,Djouadi:1999ht}, are very large and cannot be
neglected.
Such effects  are essentially independent of the Higgs mass but exhibit
a $\tan\beta $ dependence. In the low  $\tan\beta $ region, $K$ 
can be approximated by 2 while for large  $\tan\beta $ its value
gets closer to unity \cite{Baglio:2010ae}. 
In our study we have taken $K$ to be constant for fixed $\tan \beta$ in the considered range of Higgs masses.
\subsection{Indirect constraints}
As explained in the introduction, indirect searches of new physics
in low-energy precision experiments play a very important role in
Higgs boson searches. The main players in this game are $b\to s\gamma$
and $B_{s}\to\mu^{+}\mu^{-}$.
\subsubsection{$b\rightarrow s\gamma$ decay.}
Following references \cite{Degrassi:2000qf,Misiak:2006zs,Lunghi:2006hc,Gomez:2006uv}, 
the branching ratio of the decay given in terms of the
Wilson coefficients can be written as:
\begin{equation}
\mbox{BR}(B\rightarrow X_{s}\gamma)\simeq\left[a~+a_{77}~\delta\mathcal{C}_{7}^{2}+a_{88}~\delta\mathcal{C}_{8}^{2}+\mbox{Re}\left[a_{7}~\delta\mathcal{C}_{7}\right]+\mbox{Re}\left[a_{8}~\delta\mathcal{C}_{8}\right]+\mbox{Re}\left[a_{78}~\delta\mathcal{C}_{7}\delta\mathcal{C}{}_{8}^{*}\right]\right]\label{eq:5.1-1}
\end{equation}
where $a\sim3.0\times10^{-4}$, $a_{77}\sim4.7\times10^{-4}$, $a_{88}\sim0.8\times10^{-4}$,
$a_{7}\sim\left(-7.2+0.6\, i\right)\times10^{-4}$, $a_{8}\sim\left(-2.2-0.6\, i\right)\times10^{-4}$
and $a_{78}\sim\left(2.5-0.9\, i\right)\times10^{-4}$ and the main
contributions to the Wilson coefficients, beyond the $W$--boson contribution,
are chargino and charged-Higgs contributions, $\delta\mathcal{C}_{7,8}=\mathcal{C}_{7,8}^{H^{\pm}}+\mathcal{C}_{7,8}^{\chi^{\pm}}$.

Chargino contributions are given by,
\begin{equation}
\mathcal{C}_{7,8}^{\chi^{\pm}}=\frac{1}{\cos\beta}\sum_{{\scriptstyle a=1,2}}\left\{ \frac{U_{a2}V_{a1}M_{W}}{\sqrt{2}m_{\tilde{\chi}_{a}^{\pm}}}\mathcal{F}_{7,8}\left(x_{\tilde{q}\tilde{\chi}_{a}^{\pm}},x_{\tilde{t}_{1}\tilde{\chi}_{a}^{\pm}},x_{\tilde{t}_{2}\tilde{\chi}_{a}^{\pm}}\right)+\frac{U_{a2}V_{a2}\overline{m}_{t}}{2m_{\tilde{\chi}_{a}^{\pm}}\sin\beta}\mathcal{G}_{7,8}\left(x_{\tilde{t}_{1}\tilde{\chi}_{a}^{\pm}},x_{\tilde{t}_{2}\tilde{\chi}_{a}^{\pm}}\right)\right\} 
\end{equation}
where $x_{\alpha\beta}=m_{\alpha}^{2}/m_{\beta}^{2}$ and the functions
$\mathcal{F}_{7,8}(x,y,z)=f_{7,8}^{(3)}\left(x\right)-\left|\mathcal{R}_{11}^{\tilde{t}}\right|^{2}f_{7,8}^{(3)}\left(y\right)-\left|\mathcal{R}_{21}^{\tilde{t}}\right|^{2}f_{7,8}^{(3)}\left(z\right)$
and  $\mathcal{G}_{7,8}(x,y)=\mathcal{R}_{11}^{\tilde{t}}\mathcal{R}_{12}^{*\tilde{t}}f_{7,8}^{(3)}\left(x\right)-\mathcal{R}_{21}^{\tilde{t}}\mathcal{R}_{22}^{*\tilde{t}}f_{7,8}^{(3)}\left(y\right)$
with $f_{7,8}^{\left(3\right)}(x)$, 
\begin{equation}
f_{7}^{(3)}\left(x\right)=\frac{5-7x}{6\left(x-1\right)^{2}}+\frac{x\left(3x-2\right)}{3\left(x-1\right)^{2}}\ln x;\quad f_{8}^{(3)}\left(x\right)=\frac{1+x}{2\left(x-1\right)^{2}}-\frac{x}{\left(x-1\right)^{3}}\ln x;
\end{equation}
Now, using the expansion in Appendix~\ref{App:expand}, we can see that the dominants
terms in $\tan\beta$ are:
\begin{eqnarray}
\mathcal{C}_{7,8}^{\chi^{\pm}} &\simeq  M_{W}^{2}\frac{\mu M_{2}\tan\beta}{m_{\tilde{\chi}_{1}^{\pm}}^{2}-m_{\tilde{\chi}_{2}^{\pm}}^{2}}\left(\frac{f_{7,8}^{(3)}\left(x_{\tilde{q}\tilde{\chi}_{1}^{\pm}}\right)-f_{7,8}^{(3)}\left(x_{\tilde{t}_{1}\tilde{\chi}_{1}^{\pm}}\right)}{m_{\tilde{\chi}_{1}^{\pm}}^{2}}-\frac{f_{7,8}^{(3)}\left(x_{\tilde{q}\tilde{\chi}_{2}^{\pm}}\right)-f_{7,8}^{(3)}\left(x_{\tilde{t}_{1}\tilde{\chi}_{2}^{\pm}}\right)}{m_{\tilde{\chi}_{2}^{\pm}}^{2}}\right)\qquad\qquad\label{eq:C7char}\\
 & +~~M_{W}^{2}\frac{m_{t}^{2}}{m_{\tilde{t}_{1}}^{2}-m_{\tilde{t}_{2}}^{2}}\:\frac{\mu A_{t}\tan\beta}{m_{\tilde{\chi}_{1}^{\pm}}^{2}-m_{\tilde{\chi}_{2}^{\pm}}^{2}}\left(\frac{f_{7,8}^{(3)}\left(x_{\tilde{t}_{1}\tilde{\chi}_{1}^{\pm}}\right)-f_{7,8}^{(3)}\left(x_{\tilde{t}_{2}\tilde{\chi}_{1}^{\pm}}\right)}{m_{\tilde{\chi}_{1}^{\pm}}^{2}}-\frac{f_{7,8}^{(3)}\left(x_{\tilde{t}_{1}\tilde{\chi}_{2}^{\pm}}\right)-f_{7,8}^{(3)}\left(x_{\tilde{t}_{2}\tilde{\chi}_{2}^{\pm}}\right)}{m_{\tilde{\chi}_{2}^{\pm}}^{2}}\right)\nonumber 
\end{eqnarray}
and in the limit $m_{\tilde{\chi}_{1}}\simeq M_{2}\ll m_{\tilde{\chi}_{2}}\simeq\mu$,
we have,
\begin{eqnarray}
\mathcal{C}_{7,8}^{\chi^{\pm}}\simeq&-&\frac{M_{2}}{\mu}\tan\beta\frac{M_{W}^{2}}{M_{2}^{2}}\left(f_{7}^{(3)}\left(x_{\tilde{q}\tilde{\chi}_{1}^{\pm}}\right)-f_{7}^{(3)}\left(x_{\tilde{t}_{1}\tilde{\chi}_{1}^{\pm}}\right)\right) \\&-&\frac{A_{t}}{\mu}\tan\beta\,\frac{M_{W}^{2}}{M_{2}^{2}}\frac{m_{t}^{2}}{m_{\tilde{t}_{1}}^{2}-m_{\tilde{t}_{2}}^{2}}\:\left(f_{8}^{(3)}\left(x_{\tilde{t}_{1}\tilde{\chi}_{1}^{\pm}}\right)-f_{8}^{(3)}\left(x_{\tilde{t}_{2}\tilde{\chi}_{1}^{\pm}}\right)\right)\nonumber\label{C7charlim}
\end{eqnarray}
Then, the charged-Higgs contribution, including the would-be Goldstone-boson corrections to the W-boson contribution \cite{Gomez:2006uv}, is given by,
\begin{equation}
{\cal C}_{7,8}^{H^{\pm}}=\frac{1}{3\tan^{2}\beta}f_{7,8}^{(1)}(y_{t})+\frac{f_{7,8}^{(2)}(y_{t})\, +\, \left(\Delta h_{d}/h_{d} \left( 1 + \tan\beta\right) - \delta h_{d}/h_{d} \left( 1 - \cot\beta\right)\right)\,f_{7,8}^{(2)}(x_{t}) }{1+\delta h_{d}/h_{d}+\Delta h_{d}/h_{d}\tan\beta} \label{eq:C7H}
\end{equation}
with $y_{t}=m_{t}^{2}/M_{H^{\pm}}^{2}$,  $x_{t}=m_{t}^{2}/M_{W}^{2}$ and 
\begin{eqnarray}
f_{7}^{(1)}\left(x\right)&=\frac{x\left(7-5x-8x^{2}\right)}{24\left(x-1\right)^{3}}+\frac{x^{2}\left(3x-2\right)}{4\left(x-1\right)^{4}}\ln x;\quad f_{8}^{(1)}\left(x\right)&=\frac{x\left(2+5x-x^{2}\right)}{8\left(x-1\right)^{3}}-\frac{3x^{2}}{4\left(x-1\right)^{4}}\ln x;\nonumber \\
f_{7}^{(2)}\left(x\right)&=\frac{x\left(3-5x\right)}{12\left(x-1\right)^{2}}+\frac{x\left(3x-2\right)}{6\left(x-1\right)^{3}}\ln x;\qquad\quad f_{8}^{(2)}\left(x\right)&=\frac{x\left(3-x\right)}{4\left(x-1\right)^{2}}-\frac{x}{2\left(x-1\right)^{3}}\ln x;
\end{eqnarray}

\subsubsection{$B_{s}\rightarrow\mu^{-}\mu^{+}$ decay.}
The branching ratio associated to this decay can be adequately approximated
by the following expression \cite{Buras:2002vd}:
\begin{equation}
\mbox{BR}(B_{s}\rightarrow\mu^{-}\mu^{+})=2.32\cdot10^{-6}\;\frac{\tau_{B_{s}}}{1.5ps}\left(\frac{F_{B_{s}}}{230MeV}\right)^{2}\left(\frac{\left|V_{ts}\right|}{0.04}\right)^{2}\left[\left|\tilde{c}_{S}\right|^{2}+\left|\tilde{c}_{P}+0.04(c_{A}-c'_{A})\right|^{2}\right]\label{eq:bsmumu}
\end{equation}
where the dimensionless Wilson coefficients are given by $\tilde{c}_{S}=m_{B_{s}}c_{S}$,
$\tilde{c}_{P}=m_{B_{s}}c_{P}$ and the coefficients $c_{A}$ and
$c'_{A}$ can be neglected in comparison with $c_{S}$ and $c_{P}$
since they are related with contributions from box diagrams and $Z^{0}$-penguin
diagrams. In our analysis, we use the approximate expressions for  $c_{S}$ and $c_{P}$ in Ref.~\cite{Buras:2002vd}:
\begin{equation}
c_{P}\simeq\frac{m_{\mu}\overline{m}_{t}^{2}}{4M_{W}}\,\frac{16\pi^{2}\tan^{3}\beta~\epsilon_{Y}}{\left(1+\delta h_{d}/h_{d}+\Delta h_{d}/h_{d}\tan\beta\right)\left(1+\epsilon_{0}\tan\beta\right)}\left[\frac{\left|U_{11}\right|^{2}}{m_{H_{1}}^{2}}+\frac{\left|U_{21}\right|^{2}}{m_{H_{2}}^{2}}+\frac{\left|U_{31}\right|^{2}}{m_{H_{3}}^{2}}\right]\label{eq:5.3-2}
\end{equation}
\begin{equation}
c_{P}\simeq\frac{m_{\mu}\overline{m}_{t}^{2}}{4M_{W}}\,\frac{16\pi^{2}\tan^{3}\beta~\epsilon_{Y}}{\left(1+\delta h_{d}/h_{d}+\Delta h_{d}/h_{d}\tan\beta\right)\left(1+\epsilon_{0}\tan\beta\right)}\left[\frac{\left|U_{13}\right|^{2}}{m_{H_{1}}^{2}}+\frac{\left|U_{23}\right|^{2}}{m_{H_{2}}^{2}}+\frac{\left|U_{33}\right|^{2}}{m_{H_{3}}^{2}}\right]\label{eq:5.3-3}
\end{equation}
with
\begin{eqnarray}
\epsilon_{0}  = \frac{2\alpha_{s}}{3\pi}~\mu^* m_{\tilde{g}}^* ~I\left(m_{\tilde d_1}^2,m_{\tilde d_2}^2,m_{\tilde g}^2\right) \qquad & \qquad\epsilon_{Y}  =  -\frac{1}{16\pi^{2}}~  A_{t}^* \mu^*~ I\left(m_{\tilde t_1}^{2},m_{\tilde t_2}^{2},|\mu|^2 \right).
\label{eq:5.3-4}
\end{eqnarray}
And, given that in Eq.~(\ref{eq:bsmumu}) we are including only the $\tan \beta$-enhanced Higgs contributions, in the following, we use the experimental result as a 3$\sigma$ upper limit on this contribution.
\section{Model analysis.}
\label{sec:analysis}
In the previous section we have defined the MSSM model we are going
to analyze and presented the different production mechanisms and the
main decay channels for neutral Higgses at LHC. In this section we
study, in this general MSSM scenario with the possible presence of
CP violating phases, whether it is still possible to interpret the
Higgs resonance observed at LHC with a mass of $\sim125$ GeV as the
second Higgs having a lighter Higgs below this mass and a third neutral
Higgs with a mass $m_{H_{3}}\leq200$ GeV. As we will see in the following,
the present experimental results that we use to this end are the measurement
of $pp\to H_{2}\to\gamma\gamma$, $pp\to H_{a}\to\tau\tau$ at LHC
and the indirect constraints on charged Higgs from BR($b\to s\gamma$).
We divide our analysis in two $\tan\beta$ regions: low $\tan\beta$
defined as $\tan\beta\lesssim8$ and medium-large $\tan\beta$, for
$\tan \beta \gtrsim8$. 
\subsection{Medium--large $\tan\beta$ regimen.}
Now, we take $\tan\beta\gtrsim 8$, which implies that $\sin\beta\simeq1$
and $\cos\beta\simeq(1/\tan\beta)\ll1$. We analyze the different
processes in this regime of medium--large $\tan\beta.$ First, we
analyze the model predictions for the process $pp\to H_{2}\to\gamma\gamma$
that is requested to satisfy the new experimental constraints with
a signal strength $0.75\leq\mu_{\gamma\gamma}^{\rm{LHC}}\leq1.55\,.$
Then, we analyze the constraints from $pp\to H_{a}\to\tau\tau$ and
see whether the two results can be compatible in the regime of medium--large
$\tan\beta$ for $m_{H_{2}}=125$ GeV. 
\subsubsection{Two photon cross section.}
The two photon cross section through a Higgs boson can be divided,
in the narrow-width approximation, in two parts: Higgs production cross
section and Higgs decay to the two photon final state, $\sigma_{\gamma\gamma}=\sigma(pp\to H_{2})\times\mbox{BR}(H_{2}\to\gamma\gamma)=\sigma(pp\to H_{2})\times\Gamma(H_{2}\to\gamma\gamma)/\Gamma_{H_{2}}$.
Thus we have to analyze these three elements, {\it i.e.} $\sigma(pp\to H_{2})$,
$\Gamma(H_{2}\to\gamma\gamma)$ and $\Gamma_{H_{2}}$.

In first place, we are going to analyze the decay width of the Higgs boson into
two photons in our MSSM model. As a reference value, we can compare
our prediction with the Standard Model value,
\begin{equation}
S_{H}^{\gamma}=\frac{2}{3}F_{b}^{S}\left(\tau_{Hb}\right)+\frac{8}{3}F_{t}^{S}\left(\tau_{Ht}\right)-F_{1}\left(\tau_{HW}\right)\simeq\left(-0.025+i\,0.034\right)+1.8-8.3\simeq-6.54;
\end{equation}
In the MSSM, this decay width is given by the Eq.~(\ref{eq:3.3.2-5-1})
and it has both a scalar and a pseudoscalar part, receiving each one
contributions from different virtual particles:
\begin{eqnarray}
S_{H_{2}^{0}}^{\gamma} & = & S_{H_{2}^{0},b}^{\gamma}+S_{H_{2}^{0},t}^{\gamma}+S_{H_{2}^{0},W}^{\gamma}+S_{H_{2}^{0},\tilde{b}}^{\gamma}+S_{H_{2}^{0},\tilde{t}}^{\gamma}+S_{H_{2}^{0},\tilde{\tau}}^{\gamma}+S_{H_{2}^{0},\tilde{\chi}}^{\gamma}+S_{H_{2}^{0},H^{\pm}}^{\gamma};\label{eq:4.1.1-1}\\
P_{H_{2}^{0}}^{\gamma} & = & P_{H_{2}^{0},b}^{\gamma}+P_{H_{2}^{0},t}^{\gamma}+P_{H_{2}^{0},\tilde{\chi}}^{\gamma};\label{eq:4.1.1-2}
\end{eqnarray}
Once we fix the mass of the Higgs particle, $M_{H_{2}}\simeq125$
GeV, the contributions from $W$-bosons and SM fermions are completely
fixed, at least at tree level, with the only exception of the Higgs
mixings, that we take as free, and $\tan\beta$. In the case of third
generation fermions, as we have already seen, it is very important
to take into account the non-holomorphic threshold corrections from
gluino and chargino loops to the Higgs--fermionic couplings, $(g_{f}^{S},g_{f}^{P})$
and therefore we introduce an additional dependence on sfermion masses.
Nevertheless these contributions remain very simple,
\begin{equation}
S_{H_{2}^{0},W}^{\gamma}=-g_{H_{2}WW}\: F_{1}\left(\tau_{2W}\right)=-\left(\mathcal{U}_{21}\cos\beta+\mathcal{U}_{22}\sin\beta\right)\: F_{1}\left(\tau_{2W}\right)\simeq-8.3\,\left(\mathcal{U}_{22}+\frac{\mathcal{U}_{21}}{\tan\beta}\right)\,,
\end{equation}
where we have used that $F_{1}\left(\tau_{2W}\right)=F_{1}\left(0.61\right)\simeq8$.

The top and bottom quark contributions enter both in the scalar and
pseudoscalar pieces, which are both similar. The scalar contribution,
from Eq.~(\ref{eq:3.3.2-2}) and taking into account again the $\tan\beta$
regime in consideration, is given by the following approximate expression:
\begin{eqnarray}
S_{H_{2}^{0},b+t}^{\gamma} &\simeq& \frac{1}{3}\,\left[2\,\left(\mbox{Re}\left\{ \frac{\mathcal{U}_{21}+\mathcal{U}_{22}\kappa_{d}}{1+\kappa_{d}\tan\beta}\right\} \tan\beta+\mbox{Im}\left\{ \frac{\kappa_{d}\left(\tan^{2}\beta+1\right)}{1+\kappa_{d}\tan\beta}\right\} {\cal U}_{23}\right)\, F_{b}^{S}\left(\tau_{2b}\right)\right.\nonumber\\
&&\left.~+~8\,\mathcal{U}_{22}\, F_{t}^{S}\left(\tau_{2t}\right)\right];\label{eq:4.1.1-6}
\end{eqnarray}
where $\kappa_{b}$ is a parameter associated to the finite loop-induced
threshold corrections that modify the couplings of the neutral Higgses
to the scalar and pseudoscalar fermion bilinears, as defined in Eqs.~(\ref{Eq:thresholdS},\ref{Eq:thresholdP}). These parameters
are always much lower than $1$, whereas for $m_{t}=173,1$ GeV (pole
mass) and $m_{b}=4.33$ GeV (mass at $m_{t}$ scale)  the loop functions are just about
$F_{b}^{S}\simeq-0.04+i\,0.05$ and $F_{t}^{S}\simeq0.7$. In this way,
Eq.~(\ref{eq:4.1.1-6}) can be finally approximated by:
\begin{equation}
S_{H_{2}^{0},b+t}^{\gamma}\simeq1.8~\mathcal{U}_{22} +\left(-0.025+i\,0.034\right)\left[\mbox{Re}\left\{ \frac{\tan\beta}{1+\kappa_{d}\tan\beta}\right\} \,\mathcal{U}_{21}+\,\mbox{Im}\left\{ \frac{\kappa_{d}\tan^{2}\beta}{1+\kappa_{d}\tan\beta}\right\} {\cal U}_{23}\right]\,.
\end{equation}
The first contribution beyond the Standard Model that we are going
to consider is the charged Higgs boson. As we can see from Eq.~(\ref{eq:3.3.2-2}), it only takes part in the scalar part of the decay width. Its contribution
is given by:
\begin{eqnarray}
S_{H_{2}^{0},H^{\pm}}^{\gamma} & = & -g_{H_{2}^{0}H^{\pm}}\frac{\upsilon^{2}}{2m_{H^{\pm}}^{2}}F_{0}\left(\tau_{2H^{\pm}}\right),\label{eq:4.1.1-7}
\end{eqnarray}
where the self-coupling to the second neutral Higgs can be approximated as
follows for medium-large $\tan\beta$, keeping only the leading terms
in $\cos\beta$:
\begin{eqnarray}
g_{H_{2}^{0}H^{\pm}} & \simeq & \left(2\lambda_{1}\cos\beta-\lambda_{4}\cos\beta-2\cos\beta\,\mbox{Re}\left\{ \lambda_{5}\right\} +\mbox{Re}\left\{ \lambda_{6}\right\} \right)\mathcal{U}_{21} \\
 & + & \left(\lambda_{3}+\cos\beta\,\mbox{Re}\left\{ \lambda_{6}\right\} -2\cos\beta\,\mbox{Re}\left\{ \lambda_{7}\right\} \right)\mathcal{U}_{22}+\left(2\cos\beta\,\mbox{Im}\left\{ \lambda_{5}\right\} -\mbox{Im}\left\{ \lambda_{6}\right\} \right)\mathcal{U}_{23};\nonumber\label{eq:4.1.1-8}
\end{eqnarray}
The loop function, $F_{0}\left(\tau\right)$ is quite stable for small
$\tau$, for $150\mbox{ GeV}\leq m_{H^{\pm}}\leq200\mbox{ GeV}$,
$0.17\simeq(125/300)^{2}\leq\tau_{2H^{\pm}}\leq0.097\simeq(125/400)^{2}$, 
we have $F_{0}\left(\tau_{2H^{\pm}}\right)\simeq0.34$ and then,
taking,
\begin{eqnarray}
S_{H_{2}^{0},H^{\pm}}^{\gamma} & \lesssim & -0.45\left[\left(\frac{2\lambda_{1}-\lambda_{4}-2\,\mbox{Re}\left\{ \lambda_{5}\right\} }{\tan\beta}+\mbox{Re}\left\{ \lambda_{6}\right\} \right)\mathcal{U}_{21}
\right.\nonumber \\  & + & \left.
\left(\lambda_{3}+\frac{\mbox{Re}\left\{ \lambda_{6}\right\} -2\,\mbox{Re}\left\{ \lambda_{7}\right\} }{\tan\beta}\right)\mathcal{U}_{22}+\left(\frac{2\,\mbox{Im}\left\{ \lambda_{5}\right\} }{\tan\beta}-\mbox{Im}\left\{ \lambda_{6}\right\} \right)\mathcal{U}_{23}\right]\label{eq:4.1.1-9}
\end{eqnarray}
Now, we take into account that the Higgs potential couplings $\lambda_{i}=\lambda_{i}\left(g_{,}\beta,\, M_{susy},\, A_{t},\mu\right)$,
can be safely considered $\lambda_{i}\lesssim1$. Numerically, we find
a maximum $\lambda_{i}^{max}\sim0.25$ for some of them and taking
only the couplings not suppressed by $\tan\beta$ factors, we have
$\lambda_{3}\simeq-0.074$ at tree-level with the value at one-loop
typically smaller due to the opposite sign of the fermionic corrections
and $\lambda_{6}\simeq-0.14\, e^{i\alpha}$. Thus, we can expect the charged Higgs 
contribution to be always negligible when compared to the above SM contributions, even for $m_{H^\pm} \simeq 150$~GeV, and can not modify substantially the diphoton amplitude. 

The squarks involved in the two photon decay width are the ones with large
Yukawa couplings, that is, the sbottom and the stop. The scalar contribution
of these squarks is given in Eq.~(\ref{eq:3.3.2-2}) and writing
explicitly their couplings to the Higgs, it can be expressed as follows:
\begin{eqnarray}
S_{H_{2}^{0},\tilde{b}}^{\gamma} =  -\sum_{i=1,2}\frac{1}{3}g_{H_{2}\tilde{b}_{i}^{*}\tilde{b}_{i}}\frac{v^{2}}{2m_{\tilde{b}_{i}}^{2}}F_{0}\left(\tau_{2\tilde{b}_{i}}\right)=-\sum_{i=1,2}\frac{v^{2}}{6m_{\tilde{b}_{i}}^{2}}\,\left(\tilde{\Gamma}^{\alpha bb}\right)_{\beta\gamma}\mathcal{U}_{2\alpha}\mathcal{R}_{\beta i}^{\tilde{b}*}\mathcal{R}_{\gamma i}^{\tilde{b}}\, F_{0}\left(\tau_{2\tilde{b}_{i}}\right)\qquad\label{eq:4.1.1-10}\\
S_{H_{2}^{0},\tilde{t}}^{\gamma}  =  -\sum_{i=1,2}\frac{4}{3}g_{H_{2}\tilde{t}_{i}^{*}\tilde{t}_{i}}\frac{v^{2}}{2m_{\tilde{t}_{i}}^{2}}F_{0}\left(\tau_{2\tilde{t}_{i}}\right)=-\sum_{i=1,2}\frac{2v^{2}}{3m_{\tilde{t}_{i}}^{2}}\,\left(\tilde{\Gamma}^{\alpha tt}\right)_{\beta\gamma}\mathcal{U}_{2\alpha}\mathcal{R}_{\beta i}^{\tilde{t}*}\mathcal{R}_{\gamma i}^{\tilde{t}}\, F_{0}\left(\tau_{2\tilde{t}_{i}}\right)~\qquad\label{eq:4.1.1-11}
\end{eqnarray}
In the sbottom contribution, we make the expansion described in Appendix~\ref{App:expand}, taking into account that the off-diagonal terms in its mass matrix
are much smaller than the diagonal ones. This approximation leads
us to the expression:
\begin{eqnarray}
S_{H_{2}^{0},\tilde{b}}^{\gamma} & \simeq & 0.12\tan^{2}\beta\,\frac{m_{b}^{2}}{m_{\tilde{b}_{1}}^{2}}\left[\frac{\mbox{Re}\left\{ A_{b}^{*}\mu\right\} }{m_{\tilde{b}_{2}}^{2}}\mathcal{U}_{21}-\frac{\mu^{2}}{m_{\tilde{b}_{2}}^{2}}\mathcal{U}_{22}+\frac{\mbox{Im}\left\{ A_{b}^{*}\mu\right\} }{m_{\tilde{b}_{2}}^{2}\tan\beta}\mathcal{U}_{23}\right]\label{eq:sbottomgl}\\
 & \simeq & 1.2\times10^{-5}\tan^{2}\beta\left(\frac{300\mbox{ GeV}}{m_{\tilde{b}_{1}}}\right)^{2}\left[\frac{\mbox{Re}\left\{ A_{b}^{*}\mu\right\} }{m_{\tilde{b}_{2}}^{2}}\mathcal{U}_{21}-\frac{\mu^{2}}{m_{\tilde{b}_{2}}^{2}}\mathcal{U}_{22}+\frac{\mbox{Im}\left\{ A_{b}^{*}\mu\right\} }{m_{\tilde{b}_{2}}^{2}\tan\beta}\mathcal{U}_{23}\right]\nonumber 
\end{eqnarray}
where we have used that $F_{0}\left(\tau_{2\tilde{b}_{i}}\right)\simeq0.34$
for both right and left-handed sbottoms. Assuming that $A_{b}/m_{\tilde{b}_{2}},\mu/m_{\tilde{b}_{2}}\simeq O(1)$,
it is clear that the sbottom contribution can be safely neglected, as even for $\tan \beta \sim 50$ would be two orders of magnitude below the top-quark contribution.
Incidentally, the stau contribution can be obtained with the replacement
$b\leftrightarrow\tau$, and we can also expect it to be negligible
for stau masses above 100 GeV, except for the very large $\tan \beta$ region\footnote{In a recent analysis on this issue \cite{Carena:2013iba}, enhancements of the diphoton decay width of order $~40\%$ could be obtained for $\tan \beta \geq 60$ and $m_{\tilde \tau}\simeq 95$~GeV.}. 

On the other hand, we have the top squark case where there are large
off-diagonal terms in the mass matrix which can not be neglected in
comparison with the diagonal ones, specially if we intend to analyze small stop masses. This does not allow us to use the
Appendix \ref{App:expand} approximation in such a straightforward way. Nevertheless, we can still expand the chargino mass-matrix, keeping the stop mixing matrices, $\cal{R}$, and we can write Eq.~(\ref{eq:4.1.1-11}) as,
\begin{eqnarray}
S_{H_{2}^{0},\tilde{t}}^{\gamma}&\simeq&0.45\,\left[\frac{m_{t}^{2}}{m_{\tilde{t}_{1}}^{2}}\left(\left|\mathcal{R}_{11}\right|^{2}+\left|\mathcal{R}_{12}\right|^{2}\right)+\frac{m_{t}^{2}}{m_{\tilde{t}_{2}}^{2}}\left(\left|\mathcal{R}_{22}\right|^{2}+\left|\mathcal{R}_{21}\right|^{2}\right)\right]\mathcal{U}_{22}~+~0.45~\left(1-\frac{m_{\tilde{t}_{1}}^{2}}{m_{\tilde{t}_{2}}^{2}}\right)\nonumber\\&&\,\left[-\mbox{Re}\left\{ \frac{\mu m_{t}}{m_{\tilde{t}_{1}}^{2}}\mathcal{R}_{11}^{*}\mathcal{R}_{21}\right\} \mathcal{U}_{21}+ \mbox{Im}\left\{ \frac{\mu m_{t}}{m_{\tilde{t}_{1}}^{2}}\mathcal{R}_{11}^{*}\mathcal{R}_{21}\right\} \mathcal{U}_{23}+\mbox{Re}\left\{ \frac{A_{t}^{*}m_{t}}{m_{\tilde{t}_{1}}^{2}}\mathcal{R}_{11}^{*}\mathcal{R}_{21}\right\} \mathcal{U}_{22}\right]\nonumber\\
\label{eq:stopgl}
\end{eqnarray}
where we take that $F_{0}\left(\tau_{2\tilde{t}_{1}}\right)\simeq F_{0}\left(\tau_{2\tilde{t}_{2}}\right)\simeq0.34$. Regarding the stop mass, the limit provided by ATLAS and CMS sets $m_{\tilde{t}}\geq650$~GeV for the general case where the lightest neutralino mass is $m_{\tilde{\chi}_{1}^{0}}\lesssim250$~GeV \cite{ATLAS-CONF-2013-024,ATLAS-CONF-2013-037,ATLAS-CONF-2013-053,PAS-SUS-13-011}. Therefore if we typically consider upper values for $A_{t}, \mu\lesssim 3m_{\widetilde{Q}_{3}}\sim3000$~GeV for $m_{\widetilde{Q}_{3}}\lesssim1000$~GeV (higher values may have naturalness and charge and color breaking problems) the size of the coefficients associated to the equation above will be  $m_{t}^{2}/m_{\tilde{t}_{2}}^{2},~m_{t}^{2}/m_{\tilde{t}_{1}}^{2}<0.1$, $A_{t}m_{t}/m_{\tilde{t}_{1}}^{2},~\mu m_{t}/m_{\tilde{t}_{1}}^{2}\lesssim 1.2$
 and taking into account that $\mathcal{R}_{11}^{*}\mathcal{R}_{21}\leq\frac{1}{2}$, $\left|\mathcal{R}_{ij}\right|^{2}\leq1$  and $(1-m_{\tilde{t}_{1}}^{2}/m_{\tilde{t}_{2}}^{2})<1$ we obtain
\begin{equation}
S_{H_{2}^{0},\tilde{t}}^{\gamma}\lesssim0.26\left[ - \mathcal{U}_{21}+1.7\,\mathcal{U}_{22}+\mathcal{U}_{23}\right]\label{eq:stopgl2}\,,
\end{equation}
and therefore typically an order of magnitude smaller than the top quark and the W-boson contribution and without $\tan \beta$ enhancement. Nevertheless, we keep this stop contribution to take into account the possibility of a light stop, $m_{\tilde t_1} \leq 650$~GeV with a small mass difference to the LSP.

Finally, the chargino contribution is given by:
\begin{eqnarray}S_{H_{2}^{0},\tilde{\chi}^{\pm}}^{\gamma}=\sqrt{2}g\underset{{\scriptstyle i=1,2}}{\sum}\mbox{Re}\left\{ V_{i1}^{*}U_{i2}^*G_{2}^{\phi_{1}}+V_{i2}^{*}U_{i1}^*G_{2}^{\phi_{2}}\right\} \frac{v}{m_{\chi_{i}^{\pm}}}F_{f}^{S}\left(\tau_{2\tilde{\chi}_{i}}\right)\,,\\
\mbox{with~}\quad G_2^{\phi_1} = \left(\mathcal{U}_{21}-i \sin\beta~ \mathcal{U}_{23}\right)\,,\qquad G_2^{\phi_2} = \left(\mathcal{U}_{22}-i \cos\beta~ \mathcal{U}_{23}\right)\,.\nonumber\label{eq:4.7-1}
\end{eqnarray}
Using again the expansion of chargino mass matrices, Appendix \ref{App:expand}, we have the expression:
\begin{equation}
S_{H_{2}^{0},\tilde{\chi}^{\pm}}^{\gamma}\simeq2.8\left[\cos\beta\frac{M_{W}^{2}}{\mu^{2}}\mathcal{U}_{21}+\frac{M_{W}^{2}}{M_2^{2}}\mathcal{U}_{22}\right]
\end{equation}
where we have supposed that $m_{\chi_{1}^{\pm}}\simeq M_{2}\ll m_{\chi_{2}^{\pm}}\simeq\mu$, $\sin\beta\simeq1$, $F_{f}^{S}\left(\tau_{H_{2}\chi_{2}^{\pm}}\right)\simeq F_{f}^{S}\left(\tau_{H_{2}\chi_{1}^{\pm}}\right)\simeq0.7$, and neglected $(F_{f}^{S}\left(\tau_{H_{2}\chi_{1}^{\pm}}\right)- F_{f}^{S}\left(\tau_{H_{2}\chi_{2}^{\pm}}\right))/(m_{\chi_{1}^{\pm}}^2-m_{\chi_{2}^{\pm}}^2)$.
If we take $M_{W}^{2}/M_{2}^{2}\lesssim0.05$ for $m_{\chi^\pm_1}<350$~GeV from LHC limits \cite{ATLAS-CONF-2013-035,PAS-SUS-12-022}, we have,
\begin{equation}
S_{H_{2}^{0},\tilde{\chi}^{\pm}}^{\gamma}\lesssim0.15\left[~\mathcal{U}_{22}+\frac{M_{2}^2}{\mu^2}~\mathcal{U}_{21}\right]
\end{equation}
and again we see we can safely neglect the chargino contribution compared to the $W$-boson, top and bottom contributions.

Therefore, in summary, we can safely neglect the charged Higgs, chargino
and sbottom contributions to the 2-photon decay width and we can approximate
the scalar amplitude by,
\begin{eqnarray}
S_{H_{2}^{0}}^{\gamma} & \simeq & \mathcal{U}_{21}\,\left(-\frac{8.3}{\tan\beta}\left(-0.025+i\,0.034\right)\,\mbox{Re}\left\{ \frac{\tan\beta}{1+\kappa_{d}\tan\beta}\right\} \right.\nonumber \\
 & &\left.\quad\qquad- 0.45\,\left(\frac{m_{\tilde{t}_{2}}^{2}}{m_{\tilde{t}_{1}}^{2}}-1\right)\mbox{Re}\left\{ \frac{\mu m_{t}\mathcal{R}_{11}^{*}\mathcal{R}_{21}}{m_{\tilde{t}_{2}}^{2}}\right\}\right)\, +\nonumber \\&&\mathcal{U}_{22}\, \left(-6.5+0.45\,\left(\frac{m_{\tilde{t}_{2}}^{2}}{m_{\tilde{t}_{1}}^{2}}-1\right)\mbox{Re}\left\{ \frac{A_{t}^{*}m_{t}\mathcal{R}_{11}^{*}\mathcal{R}_{21}}{m_{\tilde{t}_{2}}^{2}}\right\}+0.45\,\left(\frac{m_{t}^{2}\left|\mathcal{R}_{11}\right|^{2}}{m_{\tilde{t}_{1}}^{2}}+\frac{m_{t}^{2}\left|\mathcal{R}_{22}\right|^{2}}{m_{\tilde{t}_{2}}^{2}}\right)\right)+\nonumber\\
 && {\cal U}_{23}\, \left(\left(-0.025+i\,0.034\right)\,\mbox{Im}\left\{ \frac{\kappa_{d}\tan^{2}\beta}{1+\kappa_{d}\tan\beta}\right\} +0.45\,\mbox{Im}\left\{ \frac{\mu m_{t}\mathcal{R}_{11}^{*}\mathcal{R}_{21}}{m_{\tilde{t}_{2}}^{2}}\right\} \right)\,.
\end{eqnarray}
Thus, it looks very difficult to obtain a scalar amplitude to two
photons significantly larger than the SM value taking into account
that the stop contribution can be, at most, order one. The same discussion
applies to the pseudoscalar amplitude that receives only fermionic
contributions, only top and bottom are relevant and thus is much smaller than
the scalar contribution above. The possibility of large SUSY contributions, as advocated in Refs.~\cite{Carena:2011aa,Carena:2012gp,Carena:2013iba} seems closed, at least in the MSSM with $m_{H_2}\simeq 125$~GeV. In particular, large stau contributions would require $\tan \beta \geq 50$ that we show below to be incompatible with the bounds from $H_1,H_3 \to \tau \tau$.

Next, we analyze the Higgs production cross section, presented at section~\ref{sub:Higgs-production.}. At the partonic level, this cross section receives contributions from gluon fusion and $b\bar{b}$-fusion. 

The $b\bar{b}$--fusion is tree-level at the partonic level and proportional to the bottom Yukawa coupling. Considering only the main threshold corrections to the bottom couplings, we have,
\begin{eqnarray}
\hat{\sigma}_{b\bar{b}\to H_2}&\simeq&\frac{\pi}{6}~\frac{g^{2}m_{b}^{2}}{4M_{W}^{2}}\left(\frac{\tan^{2}\beta}{\left(1+\kappa_{d}\tan\beta\right)^2}\,\left(|{\cal U}_{21}|^{2}+|{\cal U}_{23}|^{2}\right)\right)\nonumber \\&\simeq&6.8\times10^{-5}\,\frac{\tan^{2}\beta}{\left(1+\kappa_{d}\tan\beta\right)^2}\,\left(|{\cal U}_{21}|^{2}+|{\cal U}_{23}|^{2}\right)\,.
\end{eqnarray} 
This dimensionless partonic cross section must be multiplied by the $b\bar{b}$  luminosity in the proton, $\tau\: d{\cal L}^{b\bar{b}}/d\tau$, for $\tau=m_{H_2}^{2}/s$. Taking $m_{H_2} = 125$~GeV and for $\sqrt{s} = 8$~TeV, we have 
$\tau\: d{\cal L}^{b\bar{b}}/d\tau \simeq 2300$~pb from the MSTW2008 parton distributions at LO. Thus, the  $b\bar{b}$ contribution to the $pp$ cross section:
\begin{equation}
\sigma(pp\to H_{2})_{bb} \simeq 0.16\,\frac{\tan^{2}\beta}{(1+\kappa_{d}\tan\beta)^2}\,\left(|{\cal U}_{21}|^{2}+|{\cal U}_{23}|^{2}\right) \mbox{pb}\,.
\label{eq:aprbbfusion}
\end{equation}

On the other hand, gluon fusion cross section is a loop process,
\begin{equation}
\hat{\sigma}_{gg\rightarrow H_2}^{LO} = \frac{\alpha_{s}^{2}\left(M_{H_2}\right)}{256\pi} ~\frac{m_{H_2}^2}{\upsilon^2}\left[\left|S^g_2\right|^2+[\left|P^g_2 \right|^{2}\right] \simeq 4 \times 10^{-6}  \left[\left|S^g_2\right|^2+[\left|P^g_2 \right|^{2}\right]
\end{equation}
where the scalar coupling, $S^g_2$, gets contributions from both quarks and squarks, while the pseudoscalar one, $P^g_2$, receives contributions only from quarks. With regard to the squark contributions, they can be easily obtained from Eqs.~(\ref{eq:sbottomgl},\ref{eq:stopgl}), taking into account that, for $J^\gamma_{\tilde f}=1$, $S_{2,\tilde b}^{g}=3/2~S_{2,\tilde b}^{\gamma}$ and $S_{2,\tilde t}^{g}=3/8~S_{2,\tilde t}^{\gamma}$. Therefore, it is easy to see that analogously to the photonic amplitudes, we can safely neglect the sbottom and stop contributions to gluon fusion production. Thus, the scalar and pseudoscalar contributions to gluon fusion production can be approximated by,
\begin{eqnarray}
S_{2,b+t}^{g}&\simeq&0.7 \,\mathcal{U}_{22} + \left(-0.04+i\,0.05\right)\,\left[\mbox{Re}\left\{ \frac{\tan\beta}{1+\kappa_{d}\tan\beta}\right\} \,\mathcal{U}_{21}+\,\mbox{Im}\left\{ \frac{\kappa_{d}\tan^{2}\beta}{1+\kappa_{d}\tan\beta}\right\} {\cal U}_{23}\right];\nonumber\\ \\
P_{2,b+t}^{g}&\simeq& \left(-0.04+i\,0.05\right)\,\left[\mbox{Im}\left\{ \frac{\kappa_d\tan\beta}{1+\kappa_{d}\tan\beta}\right\} \,\mathcal{U}_{22}+\,\mbox{Im}\left\{ \frac{\kappa_{d}\tan^{2}\beta}{1+\kappa_{d}\tan\beta}\right\} {\cal U}_{21} \right] \nonumber\\
&+&\left[\left(-0.04+i\,0.05\right)\mbox{Re}\left\{ \frac{\tan\beta}{1+\kappa_{d}\tan\beta}\right\}-\frac{1}{\tan\beta}\right]\,\mathcal{U}_{23};
\end{eqnarray} 

\begin{figure}[t]
\noindent \begin{centering}
\includegraphics[scale=1.0]{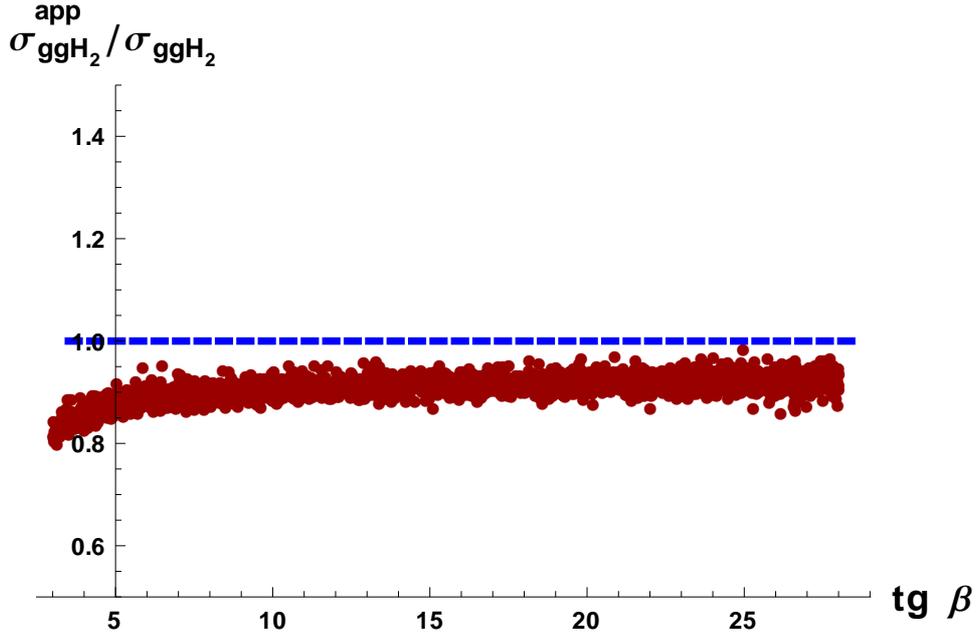}\caption{Comparison of the the approximation to $\sigma(pp\to H_{2})_{gg}$  in Eq.~(\ref{eq:aprcrosssect}) with the full result as a function of $\tan \beta$ \label{fig:siggg-H2}.}
\par\end{centering}
\end{figure}

The gluon fusion contribution to the $pp$ cross section is obtained by multiplying the gluon luminosity, $ \tau_{H_2}~d{\cal L}_{LO}^{gg}/d\tau_{H_2} \simeq 3 \times 10^6$~pb and the K-factor, which we take $K\simeq2.2$, corresponding to low $\tan \beta$. Then, with $\kappa_d$ real for simplicity, the gluon fusion contribution to $pp$ cross section would be,
\begin{eqnarray}
\label{eq:aprcrosssect}
\sigma(pp\to H_{2})_{gg}& \simeq& 27.5\, \left[\left|S^g_2\right|^2+[\left|P^g_2 \right|^{2}\right]~\mbox{pb} \simeq\left[ 13\, {\cal U}_{22}^2 -\frac{1.5 \tan\beta}{1+\kappa_{d}\tan\beta}\,{\cal U}_{21}{\cal U}_{22}\right.\\&+&\left. \frac{0.1 \tan^2\beta}{\left(1+\kappa_{d}\tan\beta\right)^2}\,{\cal U}_{21}^2 + \left(\frac{2}{\left(1+\kappa_{d}\tan\beta\right)}+\frac{0.1 \tan^2\beta}{\left(1+\kappa_{d}\tan\beta\right)^2}+\frac{27}{\tan^2\beta}\right)\,{\cal U}_{23}^2\right]~\mbox{pb} \nonumber\,.
\end{eqnarray}
This equation with the approximate values of $S^g_2, P^g_2$ is compared with the full result in Figure~\ref{fig:siggg-H2}. We can see that this approximate expression reproduces satisfactorily the gluon fusion contribution to $H_2$ production in the whole explored region.
From this equation, we see that the gluon fusion production is dominated by the top quark contribution if ${\cal U}_{21},{\cal U}_{22} = O(1)$ up to $\tan \beta \gtrsim 10$. Moreover, the SM contribution corresponds simply to take $\kappa=0$, $\tan \beta=1$, ${\cal U}_{21}={\cal U}_{22} = 1$ and ${\cal U}_{23}=0$ and therefore, we see the gluon fusion cross section will be typically smaller than the SM cross section for medium-low $\tan \beta$. Also, comparing Eqs.~(\ref{eq:aprbbfusion}) and (\ref{eq:aprcrosssect}), we see that gluon fusion still dominates over $b\bar b$--fusion except for large $\tan \beta$ or small ${\cal U}_{22}$. 

Finally, we have to check the total width, $\Gamma_{H_{2}}$. The main
decay channels for $m_{H_{2}}\simeq125$ GeV, are $H_{2}\to b\bar{b}$,
$H_{2}\to WW^{*}$ and $H_{2}\to\tau\tau$ ($H_{2}\to gg$ can of the
same order as $H_{2}\to\tau\tau$ in some cases, but, being comparatively small with respect to  $b\bar{b}$ and $WW$, it is not necessary
to consider it in the following discussion). The decay width is usually
dominated by the $b\bar{b}$-channel which can be enhanced by $\tan\beta$
factors with respect to the SM width (as the $\tau\tau$ channel).
The main contribution to the decay width to $b\bar{b}$ is captured
by the tree-level Higgs-bottom couplings, in the limit $\kappa_{d}\to0$
(although threshold corrections are important and always taken into
account in our numerical analysis),
\begin{equation}
\Gamma_{H_{2}}\simeq\frac{g^{2}m_{H_{2}}}{32\pi M_{W}^{2}}\,\left[\tan^{2}\beta\,\left(|{\cal U}_{21}|^{2}+|{\cal U}_{23}|^{2}\right)\left(3m_{b}^{2}+m_{\tau}^{2}\right)+\left(\mathcal{U}_{22}+\frac{\mathcal{U}_{21}}{\tan\beta}\right)^{2}m_{H_{2}}^{2}I_{PS}\right]\,,\label{eq:totwidth}
\end{equation}
where $I_{PS}\simeq6.7\times10^{-4}$ represents the phase space integral
in the $H_2\to W W^*$ decay width as can be found in Ref.~\cite{Lee:2003nta} for $m_{H}\simeq125$~GeV. This must
be compared with the SM decay width, which would correspond to the
usual MSSM decoupling limit if we replace $H_{1}\leftrightarrow H_{2}$
: $\tan\beta\to 1$, ${\cal U}_{21},{\cal U}_{22}\to1$ and ${\cal U}_{23}=0$.
This implies that for sizable ${\cal U}_{21},{\cal U}_{23}>\tan^{-1}\beta$,
the total width will be much larger than the SM width. Then, taking into
account that we have shown that $\Gamma_{H_{2}\to\gamma\gamma}\simeq\Gamma_{h\to\gamma\gamma}^{SM}$ we have that, for  ${\cal U}_{22}\leq 1$, the diphoton branching ratio will be smaller than the SM one. The only way to keep a large branching ratio is to take  ${\cal U}_{21},{\cal U}_{23}\lesssim\tan^{-1}\beta$, when the total width is reduced keeping $\Gamma_{H_{2}\to\gamma\gamma}$ similar to the SM. 
On the other hand, we have seen that the $H_2$ production cross section is typically smaller than the SM unless we have ${\cal U}_{22}\simeq 1$ and $H_2$ is produced through the gluon-fusion process, or $\tan \beta \gtrsim 20$ with sizeable ${\cal U}_{21},{\cal U}_{23}$ and the production is dominated by $b\bar b$ fusion. Even for this last case,  $b\bar b$ fusion, the $\tan \beta$ enhancement of the production cross section is exactly compensated by the suppression on the $H_2 \to \gamma \gamma$ branching ratio.  For gluon fusion, there is no $\tan \beta$ enhancement and thus in both cases the  $\gamma\gamma$-production
cross section is smaller than the SM one. 
Therefore, we arrive to the conclusion that the only way to increase the $\gamma\gamma$-production
cross section to reproduce the LHC results in our scenario is to {\bf decrease
the total width by suppressing the $b$-quark and the $\tau$-lepton decay
widths}. This implies having a second Higgs, $H_{2}$, predominantly
$H_{u}^{0}$, so that we decrease the couplings associated to these
fermions and consequently increase the two photons branching ratio.
This condition means, in terms of the mixing matrix elements:
\begin{eqnarray}
\mathcal{U}_{22}\sim1, &  & \mathcal{U}_{21}\simeq\mathcal{U}_{23}\leq\frac{1}{\tan\beta}\ll\mathcal{U}_{22}\label{eq:Hbounds}
\end{eqnarray}

\subsubsection{Tau-tau cross section.}

The above analysis has led us to the conclusion that, to reproduce
the $\gamma\gamma$-production cross section, we need the second lightest
Higgs to be almost purely up type. As a consequence, $H_{2}$ nearly
decouples from tau fermions and then it is unavoidable that the other
neutral Higgses inherit large down-type components, increasing thus
their decays into two $\tau$-fermions. Once more, to compute the
$\tau\tau$-production cross section through a Higgs, we must compute
$\sigma(pp\to H_{i})$, $\Gamma(H_{i}\to\gamma\gamma)$
and $\Gamma_{H_{i}}$.

The decay width $H_{i}\to\tau\tau$ is given by the following equation:
\begin{equation}
\Gamma_{H_{a}\rightarrow \tau\tau}= \frac{g_{\tau\tau}^{2}m_{H_{a}}\beta_{\tau}}{8\pi}\left(\beta_{\tau}^{2}|g_{\tau,a}^{S}|^{2}+|g_{\tau,a}^{P}|^{2}\right)\,,\label{eq:3.3.1-2}
\end{equation}
where $\tau_{i}=m_{\tau}^{2}/m_{H_{i}}^{2}$ and $\beta_{\tau}=\sqrt{1-4\tau_{i}}$.
The values of the $\tau$ scalar and pseudoscalar couplings are given by:
\begin{equation}
g_{\tau i}^{S}\simeq\frac{\tan\beta}{1+\epsilon_{\tau}\tan\beta}~\mathcal{U}_{i1}+\frac{\epsilon_{\tau}\tan\beta}{1+\epsilon_{\tau}\tan\beta}~\mathcal{U}_{i2};\qquad g_{\tau i}^{P}\simeq-\frac{\tan\beta-\epsilon_{\tau}}{1+\epsilon_{\tau}\tan\beta}~\mathcal{U}_{i3}
\end{equation}
In this case $\epsilon_{\tau}\simeq g^{2}/16\pi^{2}~(\mu M_{1}/m_{\tilde{\tau}_{2}}^{2})\simeq 2\times 10^{-3}$,
and we are taking it real. Then, we have $\epsilon_{\tau}\simeq\epsilon_{b}/20$
 being only a sub-leading correction in this case which can be 
safely neglected. Therefore we get, for $i=1,3$,
\begin{equation}
\Gamma_{i,\tau\tau}\simeq\frac{m_{H_{i}}}{8\pi}\left(\frac{gm_{\tau}}{2M_{W}}\right)^{2}\left[\tan^{2}\beta\left(\left|\mathcal{U}_{i1}\right|^{2}+\left|\mathcal{U}_{i3}\right|^{2}\right)\right]\simeq\frac{g^{2}m_{H_{i}}m_{\tau}^{2}}{32\pi M_{W}^{2}}\tan^{2}\beta\,,\label{4.1.3-3}
\end{equation}
where we used that ${\cal U}_{22}\simeq1$ and ${\cal U}_{12},{\cal U}_{32}\ll1$. 

Now we need the production cross section for $H_{1}$ and $H_{3}$. We can use  Eqs.~(\ref{eq:aprbbfusion}) and (\ref{eq:aprcrosssect}) with the replacement  $\mathcal{U}_{2j} \to \mathcal{U}_{ij}$. Then, using $\left|\mathcal{U}_{i1}\right|^{2}+\left|\mathcal{U}_{i3}\right|^{2}\simeq1$ and  $\mathcal{U}_{i2}\simeq1/\tan\beta$, we have,
\begin{eqnarray}
\sigma(pp\to H_{i})_{gg}& \simeq& 27.5\, \left[\left|S^g_2\right|^2+[\left|P^g_2 \right|^{2}\right]~\mbox{pb} \simeq\left[ 13\, {\cal U}_{i2}^2 -\frac{1.5 \tan\beta}{1+\kappa_{d}\tan\beta}\,{\cal U}_{i1}{\cal U}_{i2}\right.\\&+&\left. \frac{0.1 \tan^2\beta}{\left(1+\kappa_{d}\tan\beta\right)^2}\,{\cal U}_{i1}^2 + \left(\frac{2}{\left(1+\kappa_{d}\tan\beta\right)}+\frac{0.1 \tan^2\beta}{\left(1+\kappa_{d}\tan\beta\right)^2}+\frac{27.5}{\tan^2\beta}\right)\,{\cal U}_{i3}^2\right]~\mbox{pb}\ \nonumber\\&\simeq& \left[\frac{0.1 \tan^2\beta}{\left(1+\kappa_{d}\tan\beta\right)^2} + \frac{13 + 27.5\, {\cal U}_{i3}^2}{\tan^2 \beta} +\frac{ 2\,{\cal U}_{i3}^2 - 1.5\,{\cal U}_{i1} }{1+\kappa_{d}\tan\beta}\right] ~\mbox{pb}\,, \nonumber\\
\label{eq:aprcrosssecHi}
\sigma(pp\to H_{i})_{bb} &\simeq& 0.16\,\frac{\tan^{2}\beta}{(1+\kappa_{d}\tan\beta)^2}\,\left(|{\cal U}_{i1}|^{2}+|{\cal U}_{i3}|^{2}\right) \mbox{pb} ~\simeq~  0.16\,\frac{\tan^{2}\beta}{(1+\kappa_{d}\tan\beta)^2}~\mbox{pb} \,.
\label{eq:aprbbfusHi}
\end{eqnarray}
Therefore, we see that for $\tan \beta \gtrsim 5$ in our scenario, always with ${\cal U}_{i2}\lesssim 1/\tan \beta$, the bottom contribution to gluon fusion is larger than the top contribution and only slightly smaller than the $b\bar b$--fusion. Then we approximate the total production cross section for $H_{1,3}$,
\begin{eqnarray}
\sigma(pp\to H_{i}) &\simeq& \left[0.16~\left(\frac{ \tau_{H_i}~d{\cal L}^{bb}/d\tau_{H_i}}{2300 ~\mbox{pb}}\right) +  0.11 ~\left(\frac{ \tau_{H_i}~d{\cal L}^{gg}_{LO}/d\tau_{H_i}}{3 \times 10^6 ~\mbox{pb}}\right)\right]\,\frac{\tan^{2}\beta}{(1+\kappa_{d}\tan\beta)^2}~\mbox{pb} \,.\nonumber \\
\label{eq:aprbbfusHifin}
\end{eqnarray}

The last ingredient we need is the total width of the $H_{i}$, we
can still consider that the dominant contributions will come from
$b\bar{b}$, $\tau\tau$ and $WW^{*}$ for Higgs masses below 160
GeV. For masses above 160 GeV, the width is usually dominated by real
$W$-production and $ZZ$ or $ZZ^{*}$. Therefore, below 160 GeV,
the total width can be directly read from Eq.~(\ref{eq:totwidth})
replacing $H_{2}\to H_{i}$ and the mixing ${\cal U}_{2a}\to{\cal U}_{ia}$.
For Higgs masses above 160 GeV, always below 200 GeV in our scenario, the total width will be larger than  Eq.~(\ref{eq:totwidth}) and thus taking only  $b\bar{b}$, $\tau\tau$ and $WW^{*}$ we obtain a lower limit to $\Gamma_i$. 
In the case of $H_{1}$ and $H_{3}$, we have ${\cal U}_{i2}\ll1$
and $\left|\mathcal{U}_{i1}\right|^{2}+\left|\mathcal{U}_{i3}\right|^{2}\simeq1$.
\begin{figure}[t]
\noindent \begin{centering}
\subfloat[]{\includegraphics[scale=.55]{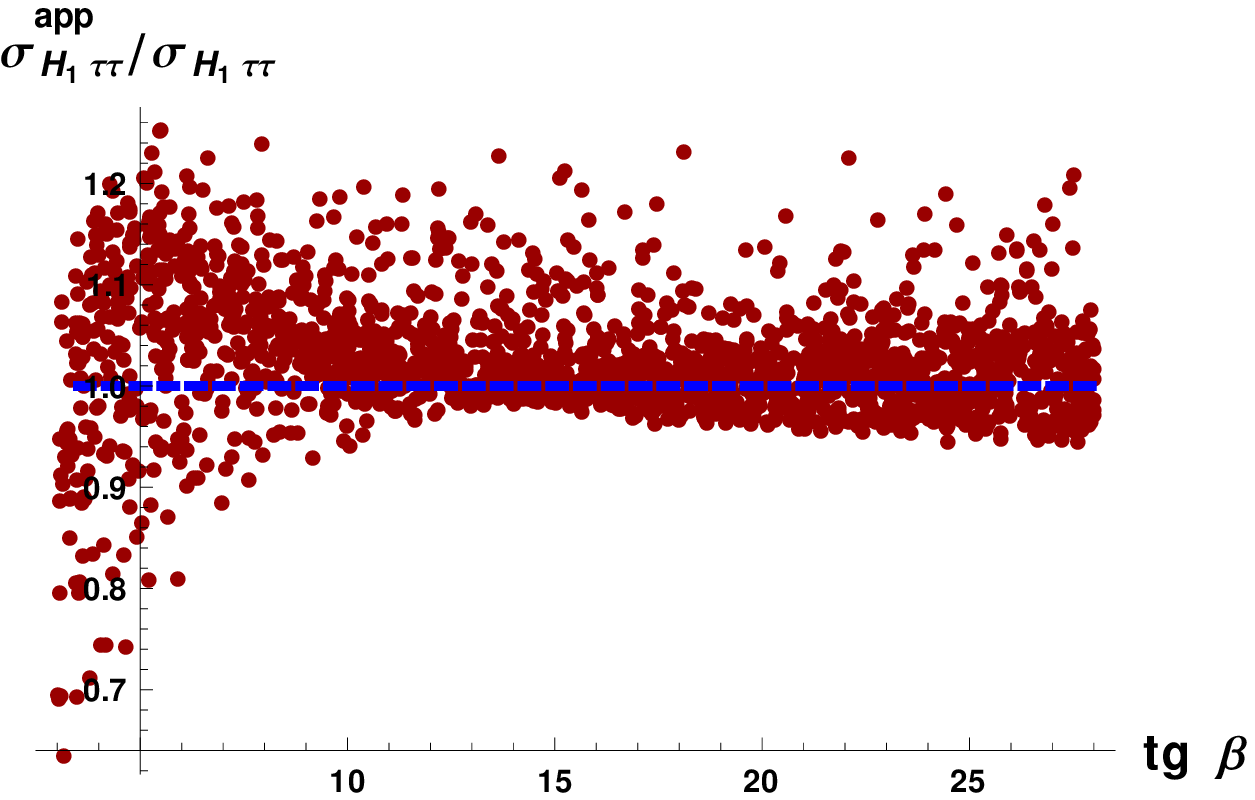}}\hspace{.2cm}\subfloat[]{\includegraphics[scale=.55]{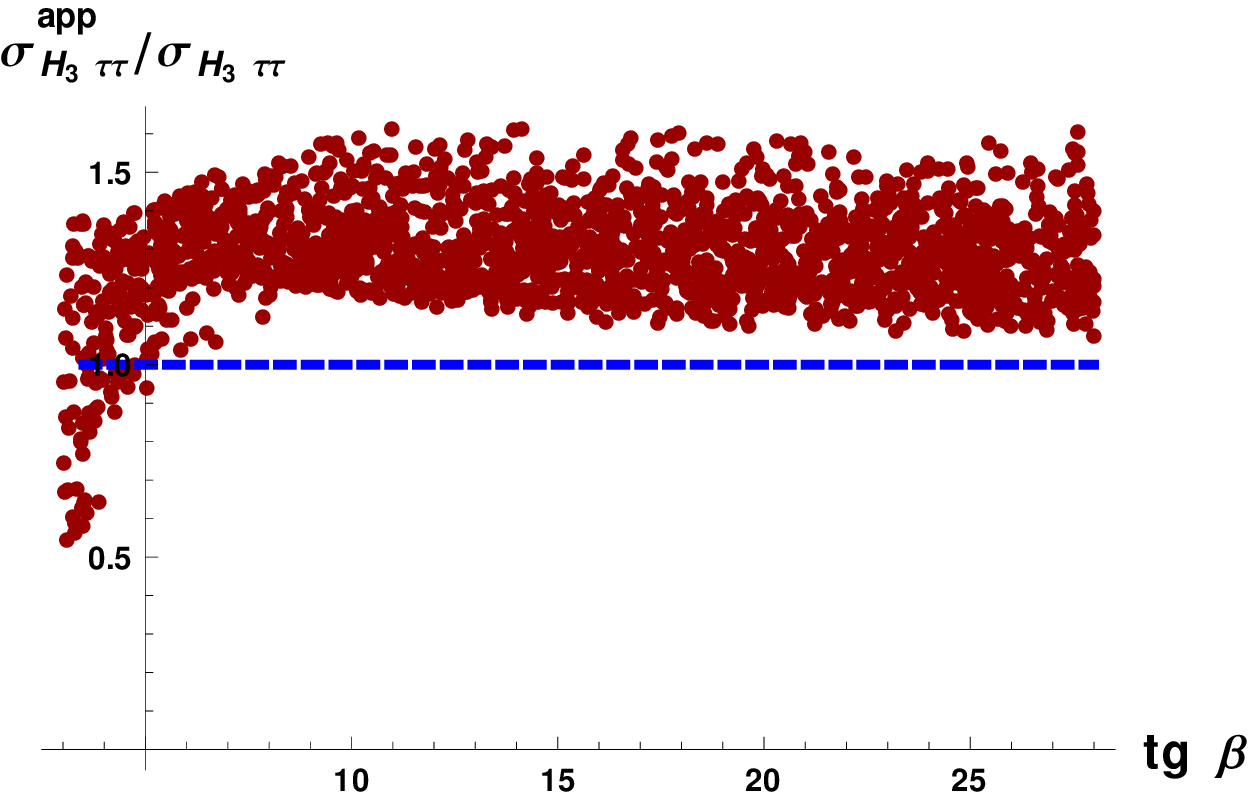}}\caption{Comparison of the the approximation to $\sigma(pp\to H_{i}\to \tau \tau)$  in Eq.~(\ref{eq:pphitauaprox}) with the full result as a function of $\tan \beta$ \label{fig:ppHitau}.}
\par\end{centering}
\end{figure} 
Then the total width is,
\begin{equation}
\Gamma_{i}\gtrsim\frac{g^{2}m_{H_{i}}}{32\pi M_{W}^{2}}\left(\frac{3m_{b}^{2}}{1+\kappa_{d}\tan\beta}+m_{\tau}^{2}\right)\tan^{2}\beta\,,
\end{equation}
And thus, the branching ratio is,
\begin{equation}
\mbox{BR}\left(H_{i}\to\tau\tau\right)\lesssim\frac{m_{\tau}^{2}\left(1+\kappa_{d}\tan\beta\right)^2}{3m_{b}^{2}+m_{\tau}^{2}\left(1+\kappa_{d}\tan\beta\right)^2}
\end{equation}
So, for the $\tau\tau$-production cross section of $H_{1}$ and $H_{3}$
we have,
\begin{eqnarray}
\label{eq:pphitauaprox}
&\sigma(pp &\overset{H_i}{\longrightarrow} \tau \tau)\lesssim\frac{\tan^{2}\beta}{\left(1+\kappa_{d}\tan\beta\right)^2}\,\frac{m_{\tau}^{2}\left(1+\kappa_{d}\tan\beta\right)^2}{3m_{b}^{2}+m_{\tau}^{2}\left(1+\kappa_{d}\tan\beta\right)^2}  \\[.2cm]
&&\qquad\qquad~\left[0.16\left(\frac{ \tau_{H_i}~d{\cal L}^{bb}/d\tau_{H_i}}{2300 ~\mbox{pb}}\right) +  0.11 \left(\frac{ \tau_{H_i}~d{\cal L}^{gg}_{LO}/d\tau_{H_i}}{3 \times 10^6 ~\mbox{pb}}\right)\right]~\mbox{pb} \nonumber\\[.2cm]
&\simeq&\frac{\tan^{2}\beta}{8.4+2\kappa_{d}\tan\beta+\kappa_{d}^2\tan^2\beta}~\left[0.16\left(\frac{ \tau_{H_i}~d{\cal L}^{bb}/d\tau_{H_i}}{2300 ~\mbox{pb}}\right) +  0.11 \left(\frac{ \tau_{H_i}~d{\cal L}^{gg}_{LO}/d\tau_{H_i}}{3 \times 10^6 ~\mbox{pb}}\right)\right]~\mbox{pb} \nonumber
\end{eqnarray}
which should be compared with the SM cross section $\sigma(pp\to H \to \tau \tau) \simeq 1.4~\mbox{pb}$ for $m_H \simeq 110$ GeV. The comparison of this approximate expression with the full result is shown in Figure~\ref{fig:ppHitau}. In fact, this approximate expression works very well for $m_{H_1}=110$~GeV and is slightly larger than the exact result for $m_{H_3}=155$~GeV. This is due to the fact that we did not include the $H_i \to W W^*$ channel in Eq.~(\ref{eq:pphitauaprox}) and this channel is important for $H_3$, which means that the approximate branching ratio is larger than one in the full expression. Nevertheless, we can safely use this expression to understand the qualitative behaviour in this process. 
 
\begin{figure}[h]
\begin{centering}
\includegraphics[scale=1.2]{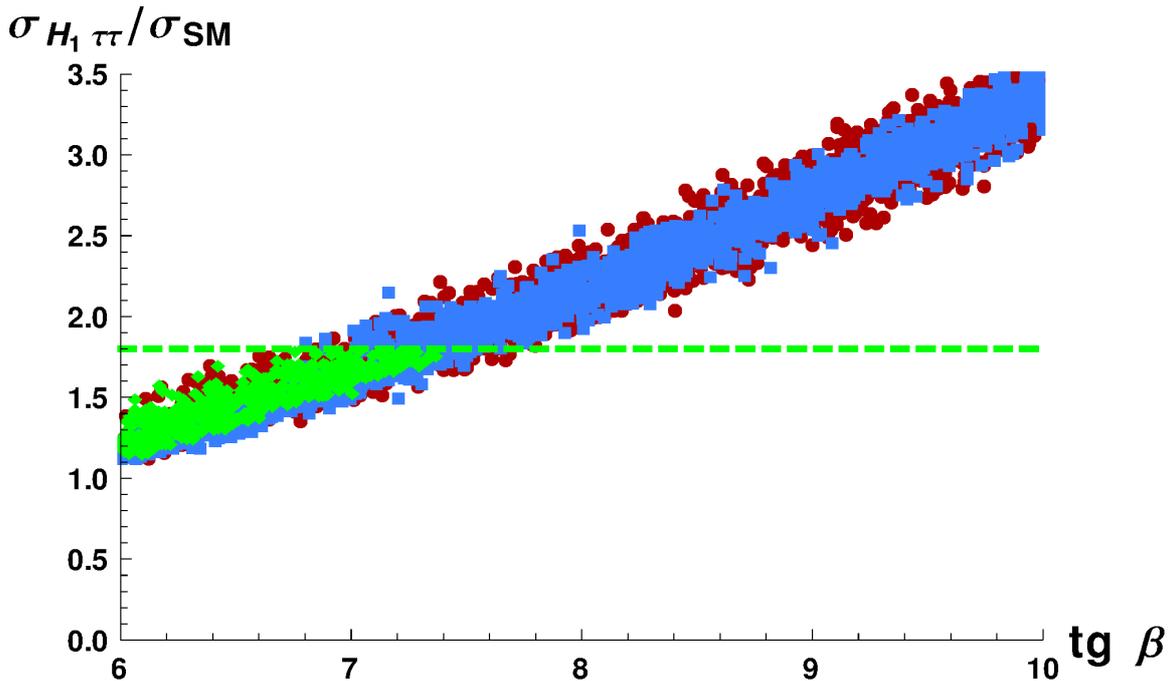}
\par\end{centering}
\caption{\label{fig:tauscatter}$\tau\tau$ production cross-section at $m_{H_1}=110$~GeV as a function of $\tan \beta$, with the CMS limit on $\tau\tau$ production in green.}
\end{figure}

Next, we combine the bounds on the two photon production cross section and the $\tau\tau$ production cross section in our model with medium-large $\tan \beta$.
In Figure \ref{fig:tauscatter} we present the $\tau\tau$ production cross sections at LHC for $m_{H_1} \simeq 110$~GeV and $m_{H_3} \simeq 160$~GeV with (squares in blue) or without (circles in red) fulfilling the requirement $0.75\leq\mu_{\gamma\gamma}^{\rm{LHC}}\leq1.55$. The green line is the CMS limit on the $\tau\tau$ production cross section for Higgs masses below 150 GeV and the green points are the points where, in addition, the $\tau\tau$ cross-section limit on the observed Higgs, $H_2$ in our scenario, at a mass $m_{H_2} \simeq 125$~GeV is also fulfilled. Even though we fixed $m_{H_1}= 110$~GeV in this plot, we have checked that the situation does not change at all for $m_{H_1}= 100$~GeV or $m_{H_1}= 120$~GeV.  
Notice that, the present constraints on heavy Higgses for $\sigma(pp\to H_3 \to \tau \tau)$ for masses $150~{\rm GeV} \leq m_{H_3} \lesssim 200~{\rm GeV}$ can only eliminate the region of $\tan \beta \gtrsim 25$, but we expect the future analysis of the stored data to reduce this parameter space significantly \cite{privateFiorini}.

Hence, we see that there are no points consistent with the LHC constraints on $\sigma(pp\to H_1 \to \tau \tau)$ for $\tan \beta \geq 7.8$ and $100~\mbox{GeV}<m_{H_1}<125$~GeV and, as we will see in the next section, all the surviving points are inconsistent with BR($B\to X_s \gamma$).

\subsection{Low $\tan\beta$ regime.}
As we have just seen, LHC constraints on $\sigma(pp\to H_1 \to \tau \tau)$ rule out the possibility of $m_{H_2}\simeq 125$~GeV for $\tan \beta \geq 7.8$, still, the situation for $\tan\beta\lesssim8$ is very different. For low $\tan \beta$, it is much easier to satisfy the constraint from the $\gamma\gamma$-signal
strength at LHC, $\mu_{\gamma\gamma}\gtrsim0.5$.

Analogously to the discussion in the case of medium-large $\tan\beta$, we can see that the $\gamma\gamma$-decay
width for low $\tan \beta$ remains of the same order as the SM one, $\Gamma_{H_{2}\to\gamma\gamma}\simeq\Gamma_{h\to\gamma\gamma}^{SM}$.
The production cross section is typically of the order of the SM one, as the
$b\bar{b}$-fusion process and the $b$-quark contribution to gluon fusion, 
being proportional to $\tan\beta$, are now smaller and the top contribution is very close to the SM for ${\cal U}_{22}\simeq O(1)$. In fact, the total decay width is still larger than the SM value if  ${\cal U}_{21,21}$ are sizeable, as the $b\bar b$ and $\tau \tau$ widths are  enhanced by $\tan^2 \beta$. So, the same requirements on Higgs mixings, Eq.~(\ref{eq:Hbounds}), hold true now, although are less suppressed correspondingly to the smaller $\tan \beta$ values.
On the other hand, the $\tau\tau$ production cross section through the three neutral Higgses remains an important constraint, but it is much easier to satisfy for low $\tan\beta$ values, as we can see in Fig.~\ref{fig:tauscatter}. 

However, in our scenario, we have a rather light charged Higgs, $m_{H^{\pm}}\lesssim220$~GeV, and the main constraint for $\tan\beta\lesssim8$ now comes from
the $\mbox{BR}(B\rightarrow X_{s}\gamma)$.
\subsubsection{Constraints from BR($B\to X_{s}\gamma$) }
The decay $B\to X_{s}\gamma$ is an important constraint on the presence
of light charged Higgs particles as we have in our scenario. However,
although the charged Higgs interferes always constructively with the
SM $W$-boson contribution to the Wilson coefficients, in the MSSM
this contribution can be compensated by an opposite sign contribution
from the stop-chargino loop if $\mbox{Re}\left(\mu A_{t}\right)$
is negative. The charged Higgs contribution is given by Eq.~(\ref{eq:C7H}). The size of ${\cal C}_7^{H^\pm}$ can be approximated by the dominant contribution, given by $f_{7}^{(2)}(m^2_{t}/m_{H^\pm}^2)$, 
\begin{equation}
{\cal C}_{7}^{H^{\pm}}\simeq\frac{f_{7}^{(2)}(y_{t})}{1+\delta h_{d}/h_{d}+\Delta h_{d}/h_{d}\tan\beta}\,, \label{eq:C7H2}
\end{equation} 
and for $m_{H^\pm} \in [150, 200]$~GeV we get $ f_{7}^{(2)}(y_{t})\in [-0.22,-0.18]$. Incidentally, we see that this charged Higgs contribution decreases with $\tan \beta$, and thus it is more difficult to satisfy the constraints at low $\tan \beta$ unless this contribution is compensated by a different sign contribution.
Then for the stop-chargino contribution, using Eq.~(\ref{C7charlim}), 
\begin{eqnarray}
\mathcal{C}_{7,8}^{\chi^{\pm}}&\simeq&-\frac{M_{W}^{2}}{M_{2}^2}~\frac{M_2}{\mu }~\tan\beta\left(f_{7,8}^{(3)}\left(x_{\tilde{q}\tilde{\chi}_{1}^{\pm}}\right)-f_{7,8}^{(3)}\left(x_{\tilde{t}_{1}\tilde{\chi}_{1}^{\pm}}\right)\right)\nonumber \\&-&\frac{A_{t}}{\mu}\tan\beta\,\frac{M_{W}^{2}}{M_{2}^{2}}\frac{m_{t}^{2}}{m_{\tilde{t}_{1}}^{2}-m_{\tilde{t}_{2}}^{2}}\:\left(f_{7,8}^{(3)}\left(x_{\tilde{t}_{1}\tilde{\chi}_{1}^{\pm}}\right)-f_{7,8}^{(3)}\left(x_{\tilde{t}_{2}\tilde{\chi}_{1}^{\pm}}\right)\right)
\end{eqnarray}
Taking now $f_{7}^{(3)}\left(x\simeq 1\right)\simeq 0.44$, and therefore, with the limits on stop and chargino masses, $m_{\tilde t_1} \geq 650$~GeV and $m_{\chi^\pm} \geq 350$~GeV, we estimate $\mathcal{C}_{7}^{\chi^{\pm}}\simeq 0.02~ M_2/\mu~ \tan \beta \ll{\cal C}_{7,8}^{H^{\pm}}$. Thus it looks very difficult to compensate the charged Higgs contribution for low $\tan\beta$ and this is confirmed in the numerical analysis.
\begin{figure}[h]
\begin{centering}
\includegraphics[scale=1.2]{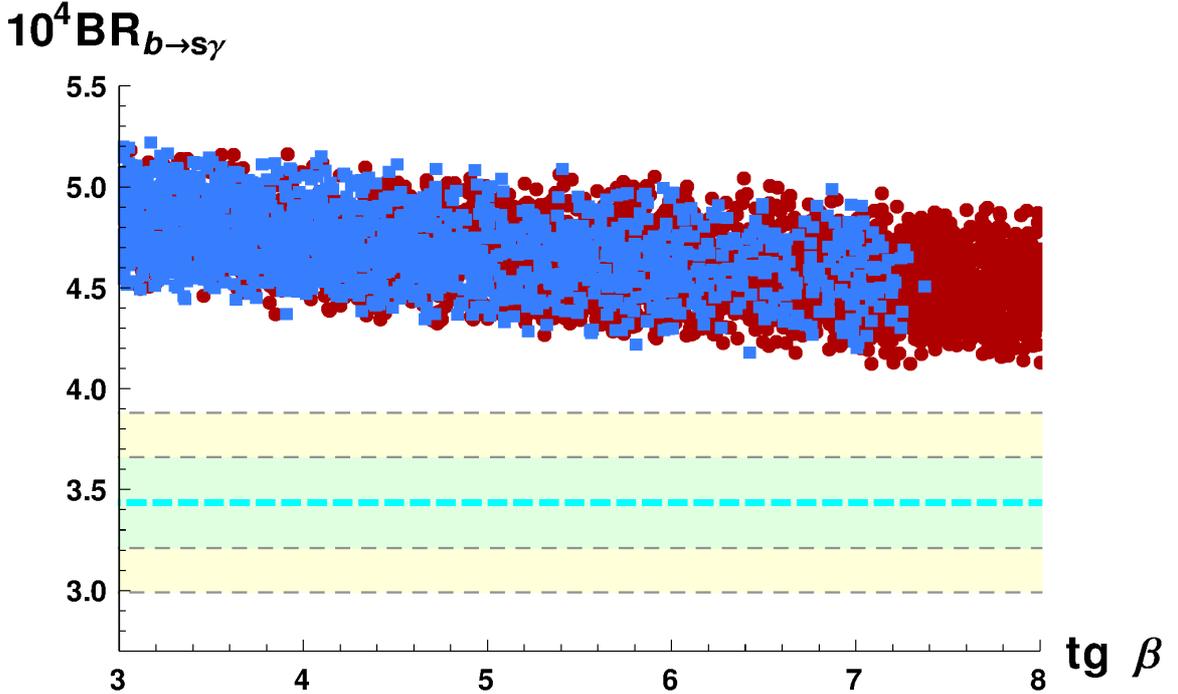}
\par\end{centering}
\caption{\label{fig:bsgammares.}Branching ratio of the $B\to X_s \gamma$ decay as a function of $\tan \beta$. Blue squares fulfil the  $\mu^{\rm LHC}_{\gamma \gamma}$ and $\sigma_{H_i\tau\tau}/\sigma_{\rm SM}$ constraints, as explained in the text. Green and yellow regions are the one and two-$\sigma$ experimentally allowed regions.}
\end{figure} 
In Figure~\ref{fig:bsgammares.}, we present the obtained BR($B\to X_s \gamma$), the blue squares fulfil the requirements of, $0.75 \leq \mu^{\rm LHC}_{\gamma\gamma}\leq 1.55$, $\sigma_{H_1\tau\tau}/\sigma_{\rm SM}\leq 1.8$ and  $\sigma_{H_2\tau\tau}/\sigma_{\rm SM}\leq 1.8$ while the red dots violate some of these requirements. The experimentally allowed region at the  one-$\sigma$ and two-$\sigma$ level is shown in green and yellow respectively \footnote{Even allowing a three-$\sigma$ range, we find no allowed points  when $m_{\tilde t_1} \geq 650$~GeV and $m_{\chi^\pm} \geq 350$~GeV}. In passing, please note that the reduction of the BR with $\tan \beta$ is mainly due to the reduction of the charged Higgs contribution, as shown in Eq.~(\ref{eq:C7H2}), and not to the negative interference with the chargino diagram.

\begin{figure}[h]
\begin{centering}
\includegraphics[scale=1.2]{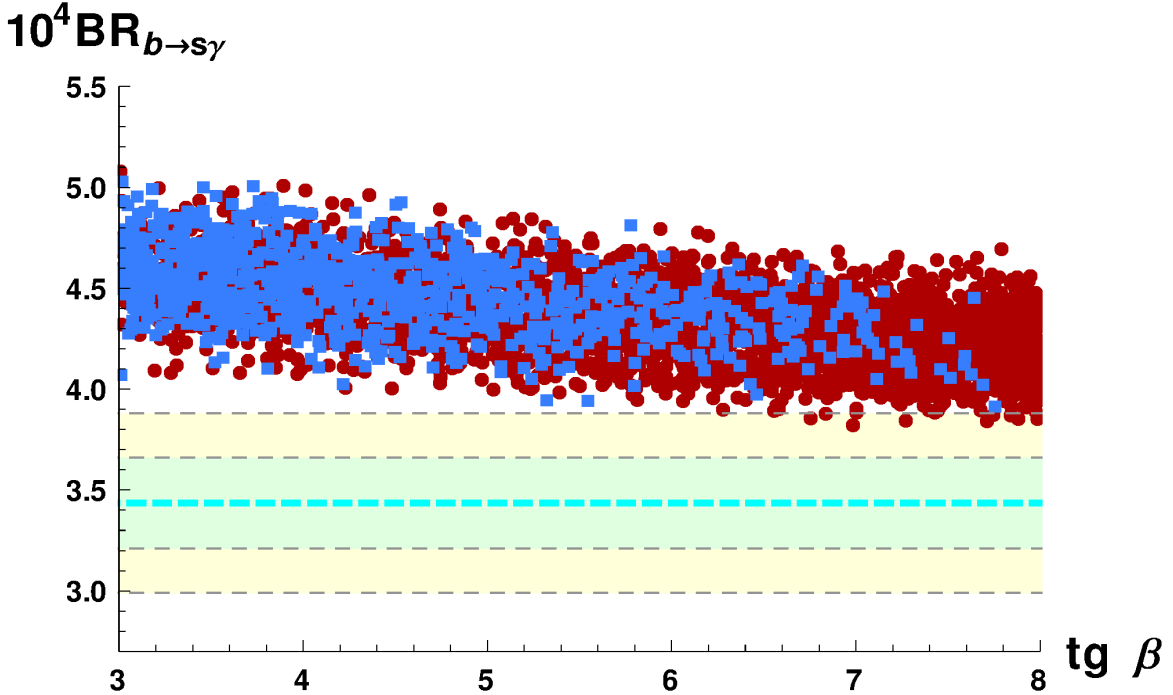}
\par\end{centering}
\caption{\label{fig:bsgammalstop}Branching ratio of the $B\to X_s \gamma$ decay as a function of $\tan \beta$, for $m_{\tilde t_1}\leq 650$ and  $m_{\chi^0_1}\leq m_{\tilde t_1} \leq m_t + m_{\chi^0_1}$. The color coding is the same as in Fig.~\ref{fig:bsgammares.}}
\end{figure}

Therefore, the only remaining option is to have a light stop with a small mass difference with respect to the lightest neutralino that has escaped detection so far at LHC. To explore numerically this possibility, we select the lightest stop mass to be $m_{\chi^0_1}\leq m_{\tilde t_1} \leq m_t + m_{\chi^0_1}$. The result is shown in Fig.~\ref{fig:bsgammalstop}, where we plot again  BR($B\to X_s \gamma$) as a function of $\tan \beta$.
Now, we can see that the range of BR($B\to X_s \gamma$) for a given $\tan \beta$ has decreased, as expected, due to a possible destructive interference of the stop-chargino diagram. Nevertheless, we can see that there are no points allowed by collider constraints that reach the two-$\sigma$ allowed region\footnote{If we allowed points within a three-$\sigma$ region,  BR$(B\to X_s \gamma) \leq 4.1 \times 10^{-4}$, several points would still survive. However, for all the three-$\sigma$ allowed points we have very large $\sigma_{H_3 \tau \tau}$ and even these points will be forbidden when ATLAS analysis on heavy MSSM Higgses is updated \cite{Aad:2012yfa,privateFiorini}.}.       

As a by-product, we can already see from here that it will be very difficult, if not completely impossible, to accommodate two sizeable Higgs-like peaks in the $\gamma\gamma$ production cross section, as recently announced by the CMS collaboration \cite{CMS-PAS-HIG-13-016}, within an MSSM context. The CMS analysis of an integrated luminosity of 5.1 (19.6) fb$^{-1}$ at a center of mass energy of 7 (8) TeV reveals a 
clear excess near $m_H=136.5$~GeV, aside from the 125--126~GeV Higgs 
boson that has already been discovered, with a local significance for this 
extra peak of 2.73 $\sigma $ combining the data  from Higgs coming from 
vector-boson fusion and vector-boson associated production (each of
which shows the excess individually).

As we have shown in this work, the ~125 GeV Higgs found at the LHC ought to be
the lightest, therefore this new resonance, despite its light mass, is bounded
to be the second lightest Higgs, meaning that the third neutral Higgs (and its
charged sibling) are to be found nearby. This can be easily seen following 
our line of reasoning in section~\ref{sec:model}, where we obtain 
$m_{H_3}< 180$~GeV and $m_{H^+}< 200$~GeV. 
However, to reproduce the observed signal strength in $H_1 \longrightarrow \gamma \gamma $ of the $\sim126$~GeV peak for medium--large $\tan \beta$,
we must force all the pseudoscalar and down-type content out of the lightest 
state. In this case, we have ${\cal U}_{12} \approx 1$ and ${\cal U}_{11},{\cal U}_{13} \ll 1$, so that the the two heavier Higgses will necessarily couple, with $\tan \beta$-enhancement, to down-type fermions and the branching ratio of these Higgses to $\gamma\gamma$ will 
be brutally inhibited. At the same time, the $H_i \longrightarrow \tau \tau $
channel, for $i=2,3$  is $\propto  (U_{i1}^2 + U_{i3}^2) \approx (1 -U_{i2}^2)
 \approx U_{12}^2\simeq 1$.
Meaning that any MSSM setting would predict a 
$H_i \longrightarrow \tau \tau $ at a level that is already excluded \cite{Aad:2012mea,CMS-PAS-HIG-13-004,Aad:2012yfa}.

The only possible escape to this situation would be to stay in the (very) low $\tan \beta$ region, but then, given the low mass of the charged Higgs, the constraints from BR($B\to X_s \gamma$) eliminate completely this possibility. Therefore, we can not see any way to accommodate two Higgs peaks in the $\gamma\gamma$ spectrum with a signal strength of the order of the SM model one. Nevertheless this possibility will be fully explored in a subsequent paper \cite{WIP}. 

\section{Conclusions.}
\label{sec:conclu}
In this work we have investigated the possibility of the
Higgs found at LHC with a mass  $m_H\sim125$~GeV  not being the lightest 
but the second lightest Higgs in an MSSM context, having the actual lightest Higgs escaped detection due to its pseudoscalar and/or down-type content. In this scheme, such a content suppresses simultaneously its couplings to gauge bosons and
up-type quarks and paves the way to evade LEP constraints.

Although similar studies, with previous LHC constraints, are already
present in the literature, most of these studies proceed through giant
scans of the model's parameter space and the later analysis of the
scanning results.  Our approach in this work has been different, and
we have chosen to study analytically, with simple expressions under
reasonable approximations, three or four key phenomenological
signatures, including the two photon signal strength and the
$\tau\tau$ production cross sections at LHC and the indirect
constraints on BR$(B\to X_s \gamma)$.  To the best of our knowledge,
this is the first study carried out in this way in an MSSM context
using the LHC data. Our approach has the advantage that can rule out
the model altogether without risking having missed a region where
unexpected cancellations or combinations can take place.  

This analysis is accomplished in a completely generic MSSM, in terms of 
SUSY parameters at the electroweak scale, such that it encloses all 
possible MSSM setups. To be as
general as possible, we have allowed for the presence of CP violating
phases in the Higgs potential such that the three neutral-Higgs eigenstates become admixtures with no definite CP--parity.
Our study starts with the $\gamma \gamma$ signal observed at LHC at $m_H\simeq 125$~GeV. The experimental results show a signal slightly larger or of the order of the SM expectations, and this is a strong constraint on models with extended Higgs sectors. We have shown than in the MSSM with  $m_{H_2}\simeq 125$~GeV the width $\Gamma(H_2 \to \gamma \gamma)$ cannot be substantially modified from its SM value. On the other hand, the total width of $H_2$ tends to be significantly larger if the down-type or pseudoscalar components of $H_2$ are sizeable. 
Simply requiring that BR$(H_2 \to \gamma \gamma)$ or, more exactly, $\sigma(pp\to H_2) \times \mbox{BR}(H_2 \to \gamma \gamma)$  is not much smaller than the SM severely restricts the possible mixings in the Higgs sector and determines the bottom and $\tau$ decay rates of the three Higgses. 

Next, we have analyzed the $\tau\tau$ production cross sections for the three Higgs eigenstates, splitting the parameter space in
two regions of large and small $\tan \beta$, being the dividing line
$\tan \beta \simeq 8$. We have shown that, for large $\tan \beta$, present constraints on $\sigma (pp \to H_1 \to \tau \tau)$ forbid all points in the model parameter space irrespective of the supersymmetric mass spectrum. 

On the other hand, in the low $\tan \beta$ region, the presence of a relatively light charged Higgs, $m_{H^\pm}\lesssim 220$~GeV, provides a large charged-Higgs contribution to $\mbox{BR}(B \to X_s \gamma)$ which can not be compensated by an opposite sign chargino contribution, precisely due to the smallness of $\tan \beta$ and this eliminates completely the possibility of the observed Higgs at $M_H \simeq 125$~GeV, being the next-to-lightest Higgs in an MSSM context.

In summary, we have shown that a carefully chosen combination of three
or four experimental signatures can be enough to entirely rule out
a model without resorting to gigantic scans while simultaneously provides a much better understanding
on the physics of the model studied. The power of this technique should not
be underestimated specially when studying models with large parameter
spaces where monster scans can be quite time consuming and not
precisely enlightening. Special interest raises the case in which the Higgs 
found at the LHC is the lightest where this type of combined analysis can
close significant regions of the parameters space \cite{WIP}.

In this respect, the straightforward application of this kind of
study to the recently published CMS data with a second Higgs-like 
resonance at $\sim 136$~GeV, aside from the 
125--126~GeV Higgs, shows it is not possible to accommodate both resonances in the $\gamma\gamma$ spectrum with a signal strength of the order of the SM model one.

\section*{Acknowledgments}
The authors are grateful to Luca Fiorini, Sven Heinemeyer, Joe Lykken and Arcadi Santamaria
for useful discussions and wish to thank specially Jae Sik Lee for his help with CPsuperH. We acknowledge support from the MEC and FEDER (EC)
Grants FPA2011-23596 and the Generalitat Valenciana under grant  PROMETEOII/2013/017. G.B. acknowledges partial support from the European Union FP7 ITN INVISIBLES (Marie Curie Actions, PITN- GA-2011- 289442).
\appendix
\section{MSSM Conventions}\label{App:convention}
We follow the MSSM conventions in the classical review of Haber and Kane \cite{Haber:1984rc}, see also \cite{Chung:2003fi}. In
this section we review the mass matrices entering in our analysis, 
\subsubsection*{Charginos:}
In our convention the chargino mass matrix is,
\begin{equation}\mathcal{M}_{C}=\left(\begin{array}{cc}
M_{2} & \sqrt{2}M_{W}\sin\beta\\
\sqrt{2}M_{W}\cos\beta & \mu
\end{array}\right)
\end{equation}
 and can be diagonalized by two unitary matrices so that $U^{*}\mathcal{M}_{C} V^{\dagger}=\mbox{Diag.}\left\{ m_{\chi_{1}^{\pm}},\: m_{\chi_{2}^{\pm}}\right\}$ with $m_{\chi_{1}^{\pm}}\leq m_{\chi_{2}^{\pm}}$.  The mass eigenstates, $\chi_i^\pm$, are related to the electroweak eigenstates, $\hat \chi_{i}^\pm$,  by
\begin{equation} 
\chi_{i}^{+}=V_{ij}\hat \chi_{j}^{+}\,, \qquad\chi_{i}^-=U_{ij}\hat\chi_{j}^-\,.
\end{equation}
\subsubsection*{Sfermions:}
The squark mass matrix is given by,
\begin{equation}
\mathcal{M}_{q}^{2}=\left(\begin{array}{cc}
M_{\tilde{Q}_{3}}^{2}+m_{q}^{2}+\cos\left(2\beta\right)M_{Z}^{2}\left(R_{z}^{q}-Q_{q}\sin^{2}\theta_{W}\right) & h_{q}^{*}\upsilon_{q}\left(A_{q}^{*}-\mu T_{q}\right)/\sqrt{2}\\
\\
h_{q}\upsilon_{q}\left(A_{q}-\mu^{*}T_{q}\right)/\sqrt{2} & M_{\tilde{R}_{3}}^{2}+m_{q}^{2}+\cos\left(2\beta\right)M_{Z}^{2}Q_{q}\sin^{2}\theta_{W}
\end{array}\right)\label{eq:3.2.2-1}
\end{equation}
With $R_{z}^{t}=-R_{z}^{b}=\frac{1}{2}$,
$Q_{q}$ the quark charge, $T_{b}=\tan\beta=\frac{\upsilon_{u}}{\upsilon_{d}}=T_{t}^{-1}$
and $h_{q}$ the Yukawa coupling corresponding to the quark. This matrix is diagonalized ${\cal R}_q \mathcal{M}_{\tilde q}^2 {\cal R}_q^\dagger=\mbox{Diag.}\left\{m_{\tilde q_{1}}^2,\:m_{\tilde q_{2}}^2\right\}$

Similarly, the stau mass matrix, 
\begin{equation}
\mathcal{M}_{\tau}^{2}=\left(\begin{array}{cc}
M_{\tilde{L}_{3}}^{2}+m_{\tau}^{2}+\cos\left(2\beta\right)M_{Z}^{2}\left(\sin^{2}\theta_{W}-\frac{1}{2}\right) & h_{\tau}^{*}\upsilon_{1}\left(A_{\tau}^{*}-\mu\tan\beta\right)/\sqrt{2}\\
\\
h_{\tau}\upsilon_{1}\left(A_{\tau}-\mu^{*}\tan\beta\right)/\sqrt{2} & M_{\tilde{E}_{3}}^{2}+m_{\tau}^{2}+\cos\left(2\beta\right)M_{Z}^{2}\sin^{2}\theta_{W}
\end{array}\right)\label{eq:3.2.2-2}
\end{equation}
\section{Expansion of Hermitian matrices}\label{App:expand}
Following Refs.~\cite{Buras:1997ij,Masiero:2005ua}, we have that given a $n\times n$ hermitian matrix $A=A^{0}+A^{1}$
with $A^{0}=Diag(a_{1}^{0},...,a_{n}^{0})$ and $A^{1}$completely
off diagonal that is diagonalized by $\mathcal{U}\cdot A\cdot\mathcal{U}^{\dagger}=Diag(a_{1},...,a_{n})$,
we have a first order in $A^{1}$:
\begin{equation}
\mathcal{U}_{ki}^{*}f\left(a_{k}\right)\mathcal{U}_{kj}\simeq\delta_{ij}f(a_{i}^{0})+A_{ij}^{1}\frac{f(a_{i}^{0})-f(a_{j}^{0})}{a_{i}^{0}-a_{j}^{0}}\label{eq:A-1}
\end{equation}
We use this formula to expand the chargino Wilson coefficients, ${\cal C}_{7,8}$,
with respect to the chargino mass matrix elements. In this case we
have to be careful because the chargino mass matrix is not hermitian.
However due to the necessary chirality flip in the chargino line ${\cal C}_{7,8}$
is a function of odd powers of $M_{\chi^{+}}$ \cite{Clavelli:2000ua},
and then
\begin{equation}
\sum_{j=1}^{2}U_{j2}V_{j1}m_{\chi_{j}^{+}}A(m_{\chi_{j}^{+}}^{2})=\sum_{j,k,l=1}^{2}U_{jk}m_{\chi_{j}^{+}}V_{j1}U_{l2}A(m_{\chi_{l}^{+}}^{2})U_{lk}^{*}
\end{equation}
where we introduced $\sum_{k}U_{jk}U_{lk}^{*}=\delta_{jl}$. Then,
we obtain,
\begin{eqnarray}
\mathcal{C}_{7,8}^{\chi^{\pm}(a)} & = & \frac{1}{\cos\beta}\sum_{{\scriptstyle a=1,2}}\frac{U_{a2}V_{a1}M_{W}}{\sqrt{2}m_{\tilde{\chi}_{a}^{\pm}}}\mathcal{F}_{7,8}\left(x_{\tilde{q}\tilde{\chi}_{a}^{\pm}},x_{\tilde{t}_{1}\tilde{\chi}_{a}^{\pm}},x_{\tilde{t}_{2}\tilde{\chi}_{a}^{\pm}}\right)\label{eq:5.2-6}\\
 & \sim & \frac{M_{W}}{\sqrt{2}\cos\beta}\left[~\left(\mathcal{M}_{\chi}\right)_{21}\frac{\mathcal{F}_{7,8}\left(x_{\tilde{q}\tilde{\chi}_{2}^{\pm}},x_{\tilde{t}_{1}\tilde{\chi}_{2}^{\pm}},x_{\tilde{t}_{2}\tilde{\chi}_{2}^{\pm}}\right)}{m_{\tilde{\chi}_{2}^{\pm}}^{2}}\right.\nonumber \\
 & + & \left.\left(\mathcal{M}_{\chi}\right)_{11}\left(\mathcal{M}_{\chi}\mathcal{M}_{\chi}^{\dagger}\right)_{21}\frac{m_{\tilde{\chi}_{1}^{\pm}}^{2}\mathcal{F}_{7,8}\left(x_{\tilde{q}\tilde{\chi}_{2}^{\pm}},x_{\tilde{t}_{1}\tilde{\chi}_{2}^{\pm}},x_{\tilde{t}_{2}\tilde{\chi}_{2}^{\pm}}\right)-m_{\tilde{\chi}_{2}^{\pm}}^{2}\mathcal{F}_{7,8}\left(x_{\tilde{q}\tilde{\chi}_{1}^{\pm}},x_{\tilde{t}_{1}\tilde{\chi}_{1}^{\pm}},x_{\tilde{t}_{2}\tilde{\chi}_{1}^{\pm}}\right)}{m_{\tilde{\chi}_{1}^{\pm}}^{2}m_{\tilde{\chi}_{2}^{\pm}}^{2}\left(m_{\tilde{\chi}_{2}^{\pm}}^{2}-m_{\tilde{\chi}_{1}^{\pm}}^{2}\right)}\right];\nonumber\label{eq:5.2-7}\end{eqnarray}\begin{eqnarray}
\mathcal{C}_{7,8}^{\chi^{\pm}(b)} & = & \frac{1}{\cos\beta}\sum_{{\scriptstyle a=1,2}}\frac{U_{a2}V_{a2}\overline{m}_{t}}{2m_{\tilde{\chi}_{a}^{\pm}}\sin\beta}\mathcal{G}_{7,8}\left(x_{\tilde{t}_{1}\tilde{\chi}_{a}^{\pm}},x_{\tilde{t}_{2}\tilde{\chi}_{a}^{\pm}}\right)\label{eq:5.2-8}\\
 & \sim & \frac{\overline{m}_{t}}{2\cos\beta\sin\beta}\left[~\left(\mathcal{M}_{\chi}\right)_{22}\frac{\mathcal{G}_{7,8}\left(x_{\tilde{q}\tilde{\chi}_{2}^{\pm}},x_{\tilde{t}_{1}\tilde{\chi}_{2}^{\pm}},x_{\tilde{t}_{2}\tilde{\chi}_{2}^{\pm}}\right)}{m_{\tilde{\chi}_{2}^{\pm}}^{2}}\right.\nonumber \\
 & + & \left.\left(\mathcal{M}_{\chi}\right)_{12}\left(\mathcal{M}_{\chi}\mathcal{M}_{\chi}^{\dagger}\right)_{21}\frac{m_{\tilde{\chi}_{1}^{\pm}}^{2}\mathcal{G}_{7,8}\left(x_{\tilde{q}\tilde{\chi}_{2}^{\pm}},x_{\tilde{t}_{1}\tilde{\chi}_{2}^{\pm}},x_{\tilde{t}_{2}\tilde{\chi}_{2}^{\pm}}\right)-m_{\tilde{\chi}_{2}^{\pm}}^{2}\mathcal{G}_{7,8}\left(x_{\tilde{q}\tilde{\chi}_{1}^{\pm}},x_{\tilde{t}_{1}\tilde{\chi}_{1}^{\pm}},x_{\tilde{t}_{2}\tilde{\chi}_{1}^{\pm}}\right)}{m_{\tilde{\chi}_{1}^{\pm}}^{2}m_{\tilde{\chi}_{2}^{\pm}}^{2}\left(m_{\tilde{\chi}_{2}^{\pm}}^{2}-m_{\tilde{\chi}_{1}^{\pm}}^{2}\right)}\right];\nonumber\label{eq:5.2-9}
\end{eqnarray}
and using again the same approximation we can expand the stop mixings
in the ${\cal F}_{7,8}$ and ${\cal G}_{7,8}$, we obtain:
\begin{eqnarray}
\mathcal{F}_{7,8}\left(x_{\tilde{q}\tilde{\chi}_{a}^{\pm}},x_{\tilde{t}_{1}\tilde{\chi}_{a}^{\pm}},x_{\tilde{t}_{2}\tilde{\chi}_{a}^{\pm}}\right) & \simeq & f_{7,8}^{(3)}\left(x_{\tilde{q}\tilde{\chi}_{a}^{\pm}}\right)-f_{7,8}^{(3)}\left(x_{\tilde{t}_{1}\tilde{\chi}_{a}^{\pm}}\right);\label{eq:5.2-10}\\
\mathcal{G}_{7,8}\left(x_{\tilde{t}_{1}\tilde{\chi}_{a}^{\pm}},x_{\tilde{t}_{2}\tilde{\chi}_{a}^{\pm}}\right) & \simeq & \left(\mathcal{M}_{\tilde{t}}\right)_{21}\frac{f_{7,8}^{(3)}\left(x_{\tilde{t}_{1}\tilde{\chi}_{a}^{\pm}}\right)-f_{7,8}^{(3)}\left(x_{\tilde{t}_{2}\tilde{\chi}_{a}^{\pm}}\right)}{m_{\tilde{t}_{1}}^{2}-m_{\tilde{t}_{2}}^{2}};\label{eq:5.2-11}
\end{eqnarray}
So, putting all together, we have: 
\begin{eqnarray}
\mathcal{C}_{7,8}^{\chi^{\pm}(a)} & \sim & \frac{M_{W}}{\sqrt{2}\cos\beta}\left[~\left(\mathcal{M}_{\chi}\right)_{21}\frac{f_{7,8}^{(3)}\left(x_{\tilde{q}\tilde{\chi}_{2}^{\pm}}\right)-f_{7,8}^{(3)}\left(x_{\tilde{t}_{1}\tilde{\chi}_{2}^{\pm}}\right)}{m_{\tilde{\chi}_{2}^{\pm}}^{2}}\right. \\
 & + & \left.\frac{\left(\mathcal{M}_{\chi}\right)_{11}\left(\mathcal{M}_{\chi}\mathcal{M}_{\chi}^{\dagger}\right)_{21}}{m_{\tilde{\chi}_{1}^{\pm}}^{2}-m_{\tilde{\chi}_{2}^{\pm}}^{2}}\left(\frac{f_{7,8}^{(3)}\left(x_{\tilde{q}\tilde{\chi}_{1}^{\pm}}\right)-f_{7,8}^{(3)}\left(x_{\tilde{t}_{1}\tilde{\chi}_{1}^{\pm}}\right)}{m_{\tilde{\chi}_{1}^{\pm}}^{2}}-\frac{f_{7,8}^{(3)}\left(x_{\tilde{q}\tilde{\chi}_{2}^{\pm}}\right)-f_{7,8}^{(3)}\left(x_{\tilde{t}_{1}\tilde{\chi}_{2}^{\pm}}\right)}{m_{\tilde{\chi}_{2}^{\pm}}^{2}}\right)\right];\nonumber\label{eq:5.2-12}\end{eqnarray}
\begin{eqnarray}
\mathcal{C}_{7,8}^{\chi^{\pm}(b)} & \sim & \frac{\overline{m}_{t}}{2\cos\beta\sin\beta}\left[~\left(\mathcal{M}_{\chi}\right)_{22}\frac{\left(\mathcal{M}_{\tilde{t}}\right)_{21}}{m_{\tilde{t}_{1}}^{2}-m_{\tilde{t}_{2}}^{2}}\left(\frac{f_{7,8}^{(3)}\left(x_{\tilde{t}_{1}\tilde{\chi}_{2}^{\pm}}\right)-f_{7,8}^{(3)}\left(x_{\tilde{t}_{2}\tilde{\chi}_{2}^{\pm}}\right)}{m_{\tilde{\chi}_{2}^{\pm}}^{2}}\right)\right. \\
 & + & \left.\frac{\left(\mathcal{M}_{\chi}\right)_{12}\left(\mathcal{M}_{\chi}\mathcal{M}_{\chi}^{\dagger}\right)_{21}}{m_{\tilde{\chi}_{1}^{\pm}}^{2}-m_{\tilde{\chi}_{2}^{\pm}}^{2}}\left(\frac{f_{7,8}^{(3)}\left(x_{\tilde{t}_{1}\tilde{\chi}_{1}^{\pm}}\right)-f_{7,8}^{(3)}\left(x_{\tilde{t}_{2}\tilde{\chi}_{1}^{\pm}}\right)}{m_{\tilde{\chi}_{1}^{\pm}}^{2}}-\frac{f_{7,8}^{(3)}\left(x_{\tilde{t}_{1}\tilde{\chi}_{2}^{\pm}}\right)-f_{7,8}^{(3)}\left(x_{\tilde{t}_{2}\tilde{\chi}_{2}^{\pm}}\right)}{m_{\tilde{\chi}_{2}^{\pm}}^{2}}\right) \right.\nonumber \\
&&\left. \frac{\left(\mathcal{M}_{\tilde{t}}\right)_{21}}{m_{\tilde{t}_{1}}^{2}-m_{\tilde{t}_{2}}^{2}} \right];\nonumber \label{eq:5.2-13}
\end{eqnarray}

\end{document}